\pdfoutput=1

\documentclass[10pt,letterpaper]{article}
\usepackage{amsmath,amssymb}

\usepackage{changepage}

\usepackage[utf8x]{inputenc}

\usepackage{textcomp,marvosym}

\usepackage{cite}

\usepackage{microtype}
\usepackage{array}

\makeatletter
\renewcommand{\@biblabel}[1]{\quad#1.}
\makeatother

\date{}

\usepackage{layouts}

\usepackage{hyperref}
\usepackage{dsfont}
\usepackage{algorithmic}
\usepackage{algorithm}
\usepackage{amsmath}
\usepackage{amssymb}
\usepackage{makeidx}
\usepackage{graphicx}
\usepackage{booktabs}
\usepackage{multirow}
\usepackage{subfigure}
\usepackage{tabularx}%
\usepackage{wrapfig}
\usepackage{amsthm}
\usepackage{tikz}
\usepackage{enumerate}

\hypersetup{colorlinks=True}
\hypersetup{urlcolor=blue}
\usepackage{placeins}

\newcommand*{\bv}[1]{\ensuremath \boldsymbol{#1}}

\newcommand*{\Xb}{\bv{X}}

\newcommand*{\fb}{\bv{f}}
\newcommand*{\gb}{\bv{g}}

\newcommand*{\yb}{\bv{y}}
\newcommand*{\tb}{\bv{t}}
\newcommand*{\Rb}{\bv{R}}

\newcommand*{\phb}{\bv{\phi}}%

\newtheorem{mydef}{Definition}

\title{Gaussian process emulation for discontinuous response surfaces with applications for cardiac electrophysiology models}
\author{Sanmitra Ghosh\textsuperscript{1,*},
	David J. Gavaghan\textsuperscript{1},
	Gary R. Mirams\textsuperscript{2}}

\begin{document}
\maketitle

\noindent\textbf{1} Computational Biology and Health Informatics, Department of Computer Science, University of Oxford, UK.
\\
\textbf{2} Centre for Mathematical Medicine \& Biology, School of Mathematical Sciences, University of Nottingham, UK.
\\
\bigskip

* sanmitra.ghosh@cs.ox.ac.uk

\section*{Abstract}
Mathematical models of biological systems are beginning to be used for safety-critical applications, where large numbers of repeated model evaluations are required to perform uncertainty quantification and sensitivity analysis.
Most of these models are nonlinear both in variables and parameters/inputs which has two consequences. 
First, analytic solutions are rarely available so repeated evaluation of these models by numerically solving differential equations incurs a significant computational burden. 
Second, many models undergo bifurcations in behaviour as parameters are varied.
As a result, simulation outputs often contain discontinuities as we change parameter values and move through parameter/input space. 

Statistical emulators such as Gaussian processes are frequently used to reduce the computational cost of uncertainty quantification, 
but discontinuities render a standard Gaussian process emulation approach unsuitable as these emulators assume a smooth and continuous response to changes in parameter values. 

In this article, we propose a novel two-step method for building a Gaussian Process emulator for models with discontinuous response surfaces.
We first use a Gaussian Process classifier to detect boundaries of discontinuities and then constrain the Gaussian Process emulation of the response surface within these boundaries. 
We introduce a novel `certainty metric' to guide active learning for a multi-class probabilistic classifier. 

We apply the new classifier to simulations of drug action on a cardiac electrophysiology model, to propagate our uncertainty in a drug's action through to  predictions of changes to the cardiac action potential.
The proposed two-step active learning method significantly reduces the computational cost of emulating models that undergo multiple bifurcations.

\section{Introduction}\label{Introduction}

Mathematical models are used ubiquitously to develop a mechanistic understanding of complex biological systems. 
However, the efficacy of these models in safety-critical applications depends on their ability to capture the interactions of several physical variables in detail in order to reproduce biological phenomena accurately \cite{PATHMANATHAN20154}.
These models are often defined as complex nonlinear dynamical systems of parameterised equations that can become intensive to computationally simulate.
Tasks such as uncertainty quantification and sensitivity analysis that require repeated evaluation with different parameter sets thus become computationally burdensome. 
A computationally cheaper alternative, an \textit{emulator}, that gives a close approximation to the responses (output) of these models is thus extremely useful for the above mentioned tasks (we refer the reader to previous work \cite{oakley2002bayesian,oakley1999bayesian} for a detailed introduction to this topic).  

Mathematical models are beginning to be used in pre-clinical drug safety and toxicology studies to learn about a compound's action on electrophysiology and associated risk \cite{davies2011silico,mirams2012application,davies2016recent}. 
The underlying mathematical models are dynamical systems described as coupled nonlinear ordinary or partial differential equations (ODEs/PDEs) that depict intra- or inter-cellular ionic exchanges and the state of cell electrophysiological components. 
Such simulations predict the occurrence of changes to the cellular \emph{action potential} (time trace for the cell's transmembrane voltage). 
Often drug-induced changes are summarised by their influence on action potential biomarkers. 
One such commonly used marker is the action potential duration (APD) which quantifies the time lag between depolarisation and repolarisation of membrane voltage. 
Commonly the output of a simulation is summarised by one or more such biomarkers that can be compared with experimental recordings.

Drug effects can be modelled by scaling the conductance parameter of multiple ion channels \cite{brennan2009multiscale}.
The degree of scaling depends on compound concentration, and is deduced from High Throughput Screening (HTS) of multiple ion channels, which is subject to considerable variability that should be taken into account as it has large effects on predictions of drug-induced changes to whole-cell electrophysiology \cite{elkins2013variability,mirams2014prediction}. 

This propagation of uncertainty, from experimental assay results (that form simulation inputs) through to simulation results, is computationally expensive, as repeated simulations involving numerical solution of differential equations have to be carried out for various input parameters.
Statistical emulators can be built that model the response surfaces spanned by the simulation outputs. 
Such an emulator can be trained using a small number of input-output pairs of the simulator; the input being the scalings applied to conductances and output being the action potential biomarkers such as APD.
Once trained, the emulator can then be used as a computationally cheaper alternative to predict the simulation output for a large number of drug blocks. 
In previous work \cite{mirams2014prediction} a simple emulator based on linear interpolation from a multi-dimensional look-up table was used to speed up the uncertainty quantification analysis we had previously performed using a `brute force' Monte Carlo method in our earliest work on this topic \cite{elkins2013variability}. 

\begin{figure}[thb]
  \centering
  \subfigure[]{
    \includegraphics[width=4.8in,height=2.5in]{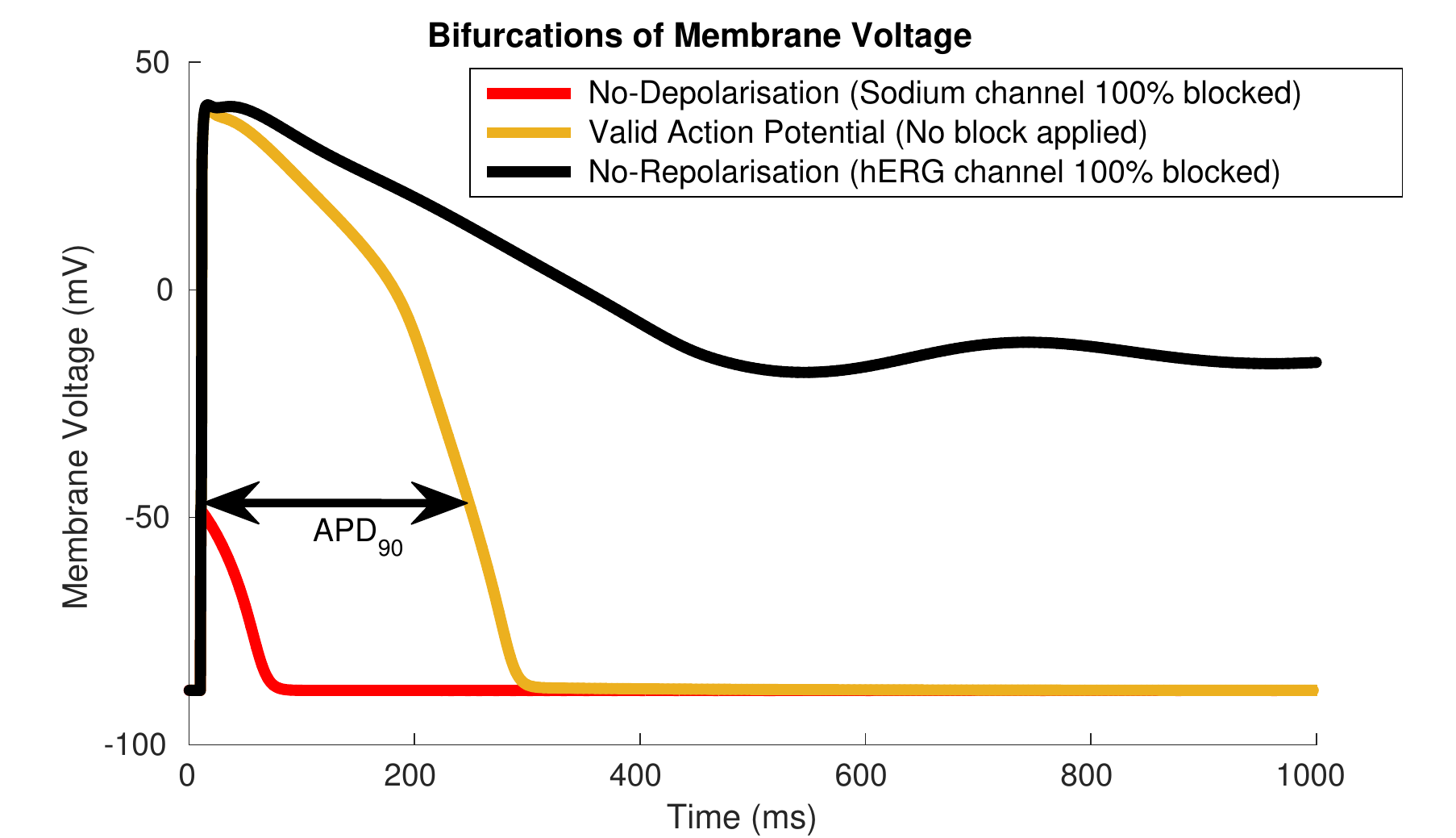}
    \label{Figure:APbifurc}
  }  
	\subfigure[]{
    \includegraphics[width=3in,height=2in]{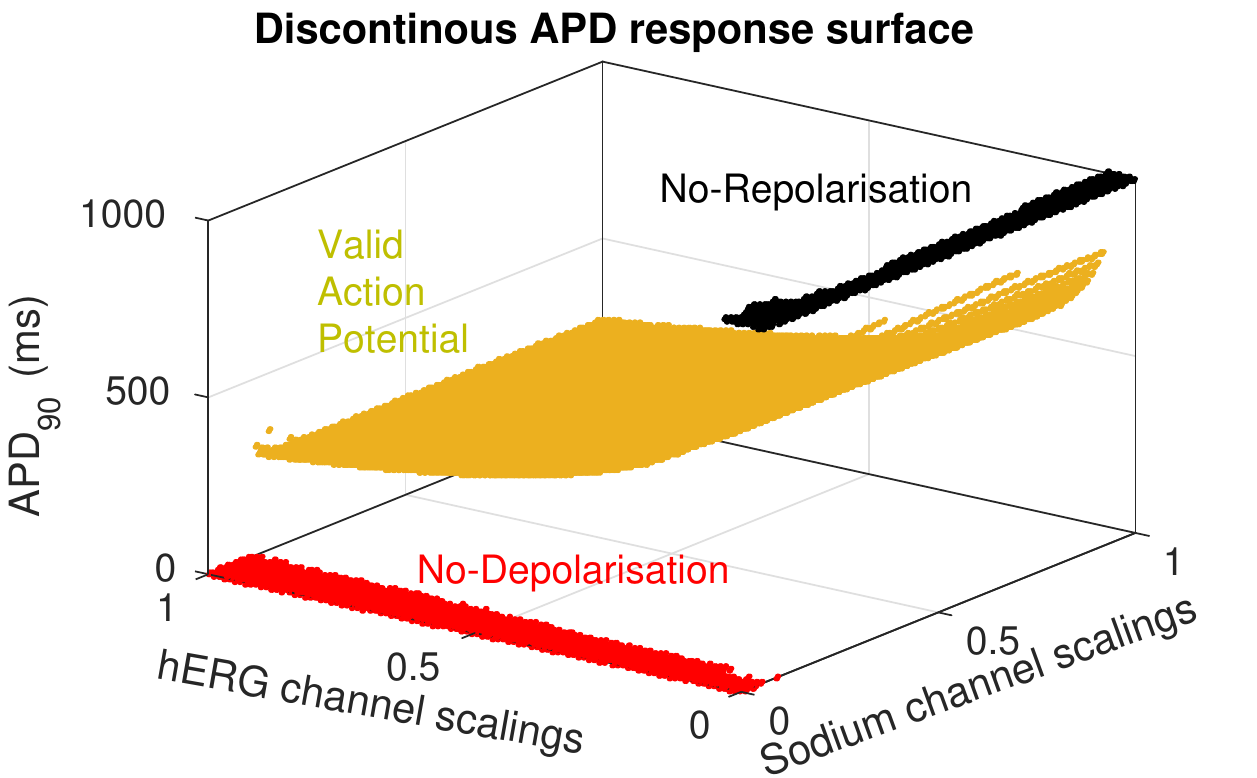}
    \label{Figure:APDsurface}
  } 
  \caption{\textbf{Bifurcation induced discontinuities in the APD biomarker response surface for the O'Hara model}. 
  a) Bifurcations in the O'Hara model dynamics \cite{o2011simulation} lead to three distinct types of behaviour, with no smooth transition in APD\textsubscript{90}.
  The black line indicates that the membrane voltage fails to return to the resting membrane potential. 
  The orange line shows a valid AP.  
  The red line shows the failure of depolarisation. 
  b) Resulting discontinuous APD response surface evaluated on a dense grid of $100 \times 100$ sample points. 
  The parameters are conductance scalings of the hERG and sodium channels. 
  The three regions of the parameter space are shaded and colour coded according to the depolarisation/repolarisation patterns as seen above.
  }
  \label{Figure:Various APs}
\end{figure}

A more efficient emulator was proposed more recently \cite{chang2015bayesian} (in terms of the number of training points required for a given accuracy) for various biomarkers obtained from simulated action potential time courses of the Luo-Rudy cardiac action potential model \cite{luo1994dynamic}.
This emulator used Gaussian Processes (GPs) to statistically model the output response surfaces of the biomarkers. 
Despite their computationally attractive properties, designing an emulator for cardiac electrophysiology models is extremely challenging since many of these models undergo bifurcations resulting in discontinuous response surfaces, as we show in Fig.~\ref{Figure:Various APs} for the widely-used O'Hara (or `ORd' model) for human ventricular action potentials \cite{o2011simulation}.

\begin{figure}[thb]
  \centering
  \includegraphics[width=\textwidth,height=3in]{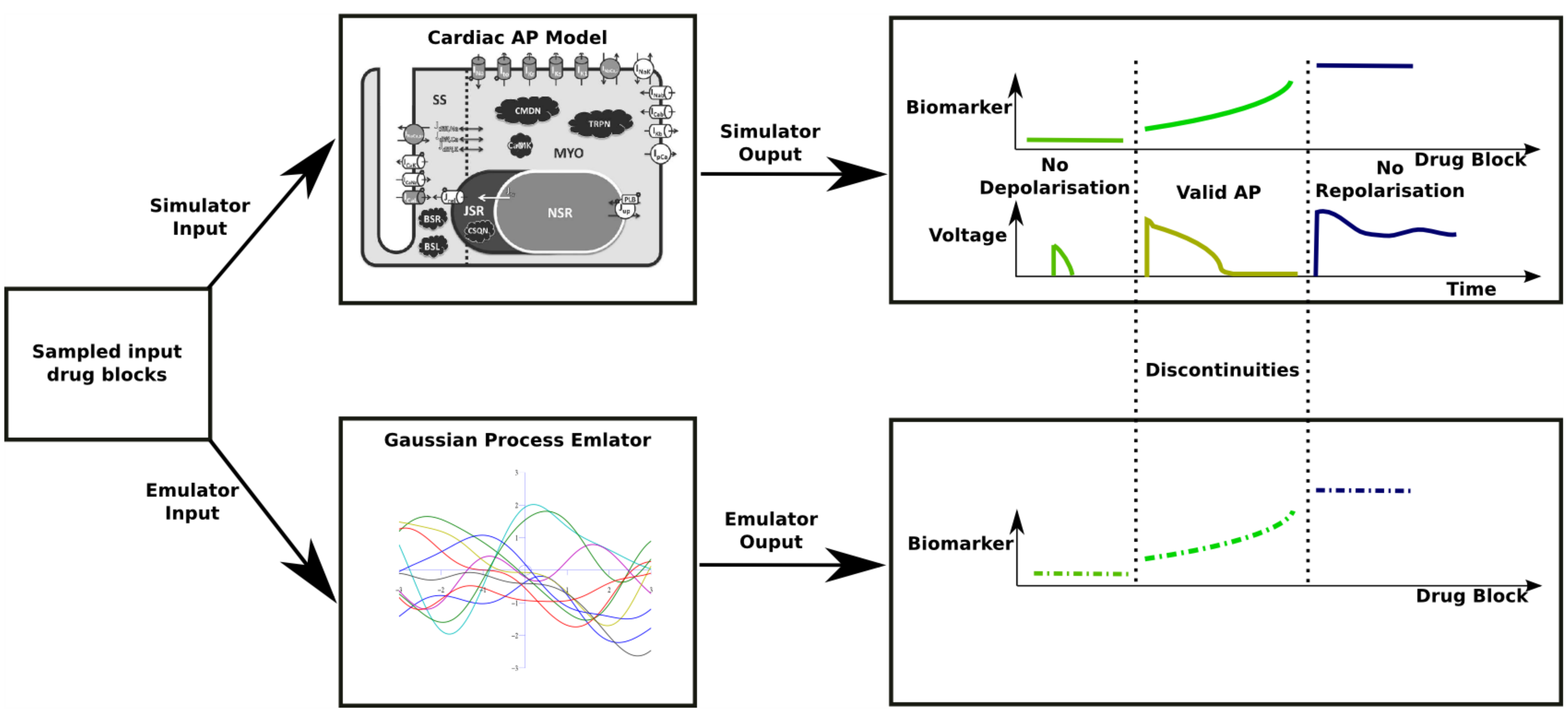}
  \caption{Comparison of the steps involved in simulation vs.\ emulation of discontinuous biomarker response surface.
  }
  \label{Figure:simulatorEmulator}
\end{figure}

In this paper we will present an emulator of APD at $90 \%$ repolarisation, APD\textsubscript{90}, in the O'Hara model, designed to work in spite of the discontinuities in the response surface.
Our proposed emulator consists of a two-step method in which we use a boundary detector to segment the response surface along the discontinuities and then apply statistical regression to emulate the responses.
We formulate the boundary detection as a classification problem.
In both these steps we use Gaussian Processes. 
The proposed emulator builds upon the work in \cite{chang2015bayesian} and \cite{Johnstone2016} with improvements to deal with these bifurcations in model behaviour. 
Fig.~\ref{Figure:simulatorEmulator} shows the main steps in both simulation and emulation of cardiac biomarkers.

\subsection{Concentration-effect curves}\label{cardiac_intro}
Our ODE system for cardiac action potentials has certain parameters that we can modify to simulate drug action. 
Typically these are the maximal current densities, or maximal conductances, of certain ion currents, denoted $G$, for various ion currents e.g.\ $G_{Kr}$, $G_{CaL}$, $G_{Na}$ etc..
Simulating the action of drug block typically means scaling a $G$ parameter by multiplying by a factor $R$ which is in the range $[0,1]$, as described below.
In what follows, the vector $\mathbf{R}$ (scaling factors $R$ for each channel $j$) then defines our parameter set or point in simulator input space.

A concentration-effect curve maps the concentration of a compound to an effect or response. 
The percentage of the peak ionic current, following a voltage step, is repeatedly recorded and the proportion that remains is the recorded effect or response $R\left(\left[C \right]  \right)$. 
Usually such curves are described by a Hill function:
\begin{equation}\label{eq:ce curve}
R\left(\left[C \right]  \right) =1-\frac{\left[C \right]^n}{\left[C \right]^n + \left[IC50 \right]^n}.
\end{equation}
This function of concentration $\left[C \right]$, has two parameters: $\left[IC50 \right]$, the
half-maximal inhibitory concentration; and the Hill coefficient $n$. 
These parameters are estimated by fitting the concentration-effect curve to screening results. 
The effect of the conductance block by a specific compound is then studied (in simulation) using a cardiac action potential (AP) model by scaling the maximal channel conductance $G$ of a particular channel $j$:
\begin{equation}\label{eq: maximal conductance}
G_{j}=G_{j,control}R_j\left([C]  \right),
\end{equation}
where $R_j\left([C]  \right) \in [0,1]$ is the conductance scaling given by the concentration-effect curve (equation~(\ref{eq:ce curve})) for ion channel $j$, and $G_{j,control}$ is the maximal conductance of that channel in control (drug free) conditions. 
This conductance scaling $R_j$ is related to the degree of ion channel block as
\begin{equation}
\%~block=100\times\big(1-R_j\left([C] \right)\big).
\end{equation}

\subsection{Handling discontinuities}\label{Aim of the study}
In Chang \emph{et al.} \cite{chang2015bayesian} the range of conductances used as input was chosen around the maximal conductances of the Luo-Rudy model in drug free (control) condition. 
Extending this range, using equation (\ref{eq: maximal conductance}), introduces other effects such as absence of depolarisation or repolarisation of the membrane voltage. 
This is caused due to the dynamical system going through bifurcations as the conductance parameters are set to values beyond a limited range around the maximal points. 
Examples of such voltage time courses are shown in Fig.~\ref{Figure:Various APs} for the O'Hara model \cite{o2011simulation}. 

For uncertainty propagation, the effect of drug block on APs \cite{mirams2014prediction} may span the entire domain of conductance scalings $R_j \in [0,1]$, where $j$ denotes any specific channel, which are applied following equation (\ref{eq: maximal conductance}). 
It is evident from Fig.~\ref{Figure:Various APs} that such a task poses a severe challenge to any emulator, as it needs to learn the location of the discontinuity of the emulated surface, as well as the value of the surface, from limited evaluations of an underlying model simulator. 

Applying GPs to model discontinuous functions is largely an open problem. 
Although many advances (see the discussion about non-stationarity in \cite{shahriari2016taking} and the references in there) have been made towards solving this problem, a common solution has not yet emerged. 
In the recent GP literature there are two specific streams of work that have been proposed for modelling non-stationary response surfaces including those with discontinuities. 
The first approach is based on designing non-stationary processes \cite{snoek2014input} whereas the other approach attempts to divide the input space into separate regions and build separate GP models for each of the segmented regions. 
Such input domain segmentation algorithms use a tree based GP model \cite{gramacy2012bayesian,assael2014heteroscedastic}. 
In such a GP model the individual nodes (leaves) of the tree are built using a smaller subset of the inputs. 
Furthermore, the model is constrained in such a way that inputs between discontinuous regions are not shared among the nodes. 
Our work is motivated by the latter approach of space partitioning which we turn to next. 

We want the emulator to return the APD\textsubscript{90} for a valid AP only (see Fig \ref{Figure:Various APs}). 
Although our emulator is general purpose and can be used with any summary statistic we concentrated on the APD\textsubscript{90} because its significance in drug induced cardiac toxicity studies -- the application for this emulator. 
Using a two-step emulator we can use a classifier to label any queried input into one of three categories according to the generated AP trace: 1) no-depolarisation, 2) a valid action potential (AP) and 3) no-repolaristaion. 
If the input falls under the second category then we can use GP regression to predict the corresponding output APD\textsubscript{90} value, and associated emulator uncertainty \cite{mirams2016}. 
Augmenting the response surface prediction step with a boundary detection we can circumvent the discontinuity while maintaining all the benefits of emulation as performed by Chang \emph{et al.}\ \cite{chang2015bayesian}. 
Using Gaussian processes for both classification and surface prediction enables us to use the uncertainty associated with the prediction to carry out sequential design of the input space. 
Thus we use `active learning' to choose training data for building the emulator which can further reduce the need for expensive space-filling designs. 
For the surface regression GP we use conditional entropy \cite{krause2008near} as a measure of uncertainty, and for the classifier GP we propose a novel metric to measure the uncertainty associated with its prediction step. 

Using a two-step emulator and carrying out sequential design for both these steps enables us to use the proposed emulator as a viable alternative at a fraction of the computational expense of solving ODEs numerically for Monte Carlo samples.

In the following section we will briefly review the fundamentals of GP regression and classification. We will then proceed to explain how we use GP regression and classification to build an emulator of the APD\textsubscript{90} response surface.

\subsection{Brief review of Gaussian processes for regression and classification} \label{sec:GP APD}

In the following sections we will review Gaussian processes for regression and classification. We will also mention briefly the approximation methods required to apply GP to larger datasets.

\subsubsection{Gaussian processes regression} \label{sec:GP Regression}
Consider the regression problem where we have the dataset $\mathcal{D}=\{(\Xb_i,y_i), i=1, \ldots, n\}$ consisting of input $\Xb_i \in \mathbb{R^D}$ and noisy scalar observations $y_i$. In the simplest case we assume that the noise is independent and Gaussian such that the \emph{latent function} $f:\mathbb{R^D} \to\mathbb{R}$ and the (possibly) noisy observations are related, using the notation in \cite{Rasmussen2006}, as
\begin{equation}\label{eq:gp regresson}
y_i=f(\Xb_i) + \epsilon_i,
\end{equation}
where $\epsilon_i \sim \mathcal{N}(0, \sigma^2_{noise})$. 
In our application the `latent function' is the simulator output that we wish to emulate. 
Also, in the context of our application we consider the noise term $\epsilon$ to be very small indeed as we are emulating the output of a deterministic ODE system, but it could represent inaccuracies introduced by numerical solution of the cardiac model. In a probabilistic framework we are interested in the probability distribution of function values $f_*$ (or the noisy $y_*$) at test locations (where we have not run the simulator but wish to infer values from the emulator) that we call $\Xb_*$.

\begin{mydef}
A Gaussian process is a collection of random variables,
any finite number of which have consistent joint Gaussian distributions
\end{mydef}

Gaussian process regression is a Bayesian approach where we place a prior over functions. 
For the regression problem we assume a priori that the function values behave according to 
\begin{equation}
p(\fb|\Xb_1,\Xb_2, \ldots, \Xb_n )=\mathcal{N}(0, K),
\end{equation}
where $\fb=( f_1, \ldots, f_n )^T$ is a vector of latent function values $f_i=f(\Xb_i)$ and $K$ is a covariance matrix whose entries are given by the covariance function $K_{i,j}=k(\Xb_i,\Xb_j;\phb)$, where $\phb$ is a vector of hyperparameters for the covariance function. 
A common example of a covariance function is the squared exponential function:
\begin{equation}
k(\Xb_i,\Xb_j)=\nu^2\exp\left( \frac{\|\Xb_i-\Xb_j\|^2}{l^2} \right),\label{eq:squarred_exp}
\end{equation}
with the hyperparameters $\phb=\{\nu,l\}$, where $\nu$ is the prior variance and $l$ is a lengthscale parameter that defines the decay rate in space of the covariance between points on the response surface. 

Inference of latent function values for test locations is carried out by first placing a joint prior on the training and test latent function values $\fb$ and $\fb_*$
\begin{equation}
\begin{bmatrix}
\fb\\
\fb_*
\end{bmatrix} \sim \mathcal{N}\left(\mathbf{0},\begin{bmatrix}\mathbf{K_{f,f}} & \mathbf{K_{f,*}} \\  \mathbf{K_{*,f}} &  \mathbf{ K_{*,*} } \end{bmatrix} \right), 
\end{equation}
where $\mathbf{K}$ is subscripted by the variables between which the covariance is computed (and we use the asterisk $*$ as shorthand for $\fb_*$). We then combine the prior with a likelihood $p(\yb|\fb)=\mathcal{N}(0,\sigma^2_{noise}\mathbb{I})$, $\mathbb{I}$ is the identity matrix, and using Bayes's rule we obtain the joint posterior
\begin{equation}
p(\fb,\fb_*|\yb)=\frac{p(\fb,\fb_*)p(\yb|\fb)}{p(\yb)},
\end{equation}
 where $\yb=\left(y_1, \ldots, y_n\right)$ is the vector of observations. Note that we have dropped the conditioning on inputs while defining the above probabilities for notational simplicity. 
 However, these probabilities defining a GP model are always conditional on the corresponding inputs. By marginalizing the training set latent function values $\fb$ we get the desired posterior function values at test locations $\Xb_*$ given by
 \begin{equation}\label{eq:gp posterior latent}
 p(\fb_*|\yb)=\int p(\fb,\fb_*|\yb)d\fb=\frac{1}{p(\yb)}\int p(\yb|\fb)p(\fb,\fb_*)d\fb.
 \end{equation}
Since both the joint GP prior and the likelihood are Gaussian we can evaluate the above integral analytically to obtain the posterior latent function at the test locations given by
\begin{equation}
p(\fb_*|\yb)=\mathcal{N}(\bf{m_f},\bf{\Sigma_f}),
\end{equation}
with the following first and second moment \cite{Rasmussen2006}:
\begin{eqnarray}\label{eq: GP post meanvar}
\begin{aligned}
&\bf{m_f}=\mathbf{K_{*,f}}(\mathbf{K_{f,f}}+\sigma^2_{noise}\mathbb{I})^{-1}\yb\\
&\bf{\Sigma_f}=\mathbf{K_{*,*}}-\mathbf{K_{*,f}}(\mathbf{K_{f,f}}+\sigma^2_{noise}\mathbb{I})^{-1}\mathbf{K_{f,*}}
\end{aligned}
\end{eqnarray}
The hyperparameters $\phb$ can be obtained as maximum likelihood estimates by maximizing the log marginal likelihood given by
\begin{equation}\label{eq:marginal likelihood}
\begin{aligned}
\log P(\yb|\phb)=-\frac{1}{2}\yb^{T}(\mathbf{K_{f,f}}+\sigma^2_{noise}\mathbb{I})^{-1}\yb-\frac{1}{2}\log \left|(\mathbf{K_{f,f}}+\sigma^2_{noise}\mathbb{I})^{-1}\right|-\frac{1}{2}\log (2\pi).
\end{aligned}
\end{equation}
We discuss how GP emulators can be refined in terms of choosing training sites to evaluate the latent function in Section~\ref{sec:Active learning for surface emulation}.

\subsubsection{Gaussian processes classification} \label{sec:GP Classification}

As mentioned previously, we will tackle discontinuities present in the simulator response surface using a boundary detector built using a classifier. 
Gaussian processes can be used for classification purposes in a discriminative probabilistic \cite{Rasmussen2006} framework. 
Thus we would use a GP classifier to detect the boundaries of discontinuities. Furthermore, we would exploit its probabilistic predictions to propagate uncertainty about the boundaries for carrying out active learning (we discuss this in section \ref{sec:Active learning}). 
Next, we briefly review the method for GP classification.

In a classification problem the input remains the same as that of regression but the outputs are discrete class occupancy labels $t_i\in \{-1,+1\}$ (`$-1$' for not in the class of interest, and `$+1$' for in the class).
We are interested in predicting the class membership for a test point $\Xb_*$. 
This is achieved using a latent function $g(\Xb_*)$ whose value is mapped to the unit interval by means of a sigmoid function $sig:\mathbb{R} \to [0,1]$ such that \cite{Rasmussen2006}
\begin{equation}
\pi:= p(t_*=+1|\Xb_*)=sig(g(\Xb_*)).
\end{equation}
The class membership probability must normalise, $\sum_{t_*}p(t_*=+1|\Xb_*)=1$,  thus we have $p(t_*=-1|\Xb_*)=1-p(t_*=+1|\Xb_*)$. 
The sigmoid function takes the form
\begin{equation}
sig(g(\Xb))=\frac{1}{1+\exp^{-g(\Xb)}}.\label{eqn:sigmoid}
\end{equation}

The likelihood of the class labels $\tb=(t_1, \ldots, t_n)$ for $n$ data points is assumed to be a Bernoulli distribution given by
\begin{equation}\label{eq: classifier likelihood}
p(\tb|\gb)=\prod_{i=1}^n p(t_i|g_i)=\prod_{i=1}^n sig(t_i|g_i), 
\end{equation} 
where we have assumed that the class labels are i.i.d. and $\gb=( g_1, \ldots, g_n )$ is the vector of latent function values.

Now just like the regression case, we can put a joint prior $p(\gb,\gb_*)$ on the training ($\gb$) and test ($\gb_*$) latent function values. This immediately enables the use of the standard GP machinery to obtain the posterior predictive distribution over the class labels
\begin{equation}\label{eq: GP classifier class}
\pi:= p(\tb_*=+1|\tb)=\int sig(\tb_*|\gb_*)p(\gb_*|\tb)d\gb_*.
\end{equation} 
Unfortunately the posterior term $p(\gb_*|\tb)$ is intractable as it involves an integration over the likelihood given by equation (\ref{eq: classifier likelihood}) with a sigmoid nonlinearity.
Approximation schemes can be used to overcome this intractability. 
While MCMC methods provide the closest approximation \cite{barber1996williams} such methods are often found to be extremely slow for a datasets of even moderate size. 
Expectation propagation (EP, \cite{minka2001expectation}) is an iterative deterministic approximation scheme that is widely used for inference in GP classifiers as it provides good accuracy and is much faster than MCMC. 
See Nickisch \& Rasmusson \cite{nickisch2008approximations} for a review and comparison of different approximations for inference in a binary GP classifier.
In EP, the individual (per data point) sigmoidal likelihood terms are approximated by un-normalised Gaussians $\xi_i(g_i)$. 
We term these local Gaussian approximations as \textit{site} functions. Thus, the likelihood $p(t_i|g_i)$ for $i$-th data point $t_i$ is approximated as
\begin{equation}
\begin{aligned}
p(t_i|g_i)&=sig(t_i|g_i),\\
&\approx \xi_i(g_i;\tilde{\mu}_i,\tilde{\sigma}_i,\tilde{Z}_i),\\
&\approx \tilde{Z}_i\mathcal{N}(g_i;\tilde{\mu}_i,\tilde{\sigma}_i),
\end{aligned}
\end{equation}
with site parameters $\{\tilde{\mu}_i,\tilde{\sigma}_i,\tilde{Z}_i\}$. For convenience we can write the product of the local approximate likelihoods as
\begin{equation}
\prod_{i=1}^n \xi_i(g_i;\tilde{\mu}_i,\tilde{\sigma}_i,\tilde{Z}_i) = \mathcal{N}(\tilde{\bv{\mu}},\tilde{\Sigma})\prod_{i=1}^n\tilde{Z}_i,
\end{equation}
where $\tilde{\bv{\mu}}=(\tilde{\mu}_1, \ldots, \tilde{\mu}_n)$ and $\tilde{\Sigma}$ is a diagonal matrix with diagonal elements $\tilde{\sigma}_i$.
The posterior latent function $p(\gb|\tb)$ is then approximated using the site functions as
\begin{equation}\label{eq:class post}
p(\gb|\tb)\approx q(\gb|\tb)=\frac{1}{Z_{EP}}p(\gb)\prod_{i=1}^n\tilde{Z}_i\mathcal{N}(g_i;\tilde{\mu}_i,\tilde{\sigma}_i)=\mathcal{N}(\bv{\mu},\Sigma),
\end{equation}
where $p(\gb|\phb)$ is the standard Gaussian prior with covariance $K$ and $Z_{EP}=p(\tb|\phb)=\int p(\tb|\gb)p(\gb|\phb)$ is the marginal likelihood. 
The posterior mean and variance is given by \cite{Rasmussen2006}
\begin{equation}
\begin{aligned}
\bv{\mu}=&\Sigma\tilde{\Sigma}^{-1}\tilde{\bv{\mu}},\\
\Sigma=&(K^{-1} + \tilde{\Sigma}^{-1})^{-1}.
\end{aligned}
\end{equation}
Note that we have made the parameter dependency explicit in the prior $p(\gb|\phb)$ while defining the marginal likelihood as in the regression case. 

The task of EP is then to find each of the site parameters iteratively so that the marginal posterior is as accurate as possible. 
To this end, we first combine the prior and the site functions into an approximate marginal distribution, also known in machine learning parlance as a \textit{cavity} distribution \cite{Rasmussen2006}: 
\begin{equation}
q_{-i}(g_i) \propto \int p(\gb)\prod_{j\neq i}\xi_j(g_j;\tilde{\mu}_j,\tilde{\sigma}_j,\tilde{Z}_j)dg_j,
\end{equation} 
where the subscript `$-i$' means ``all but the $i$-th''. 
The cavity distribution is an approximation to the marginal distribution of the latent function at the $i$-th site obtained by combining the prior $p(\gb)$ with $n-1$ (all but the $i$-th) approximate likelihood terms $\xi_i$. The simplest way to to obtain the cavity distribution is by first finding the $i$-th approximate posterior from the joint in equation (\ref{eq:class post}) and then dividing it by the $i$-th site function $\xi_i$. Thus we have the cavity distribution at the $i$-th site as \cite{Rasmussen2006}
\begin{equation}
q_{-i}(g_i)=\mathcal{N}(g_i|\mu_{-i},\sigma^2_{-i}),
\end{equation}
where $\mu_{-i}=\tilde{\sigma}^2_{-i}(\sigma^{-2}_i\mu_i - \tilde{\sigma}^{-2}_i\tilde{\mu}_i)$ and $\sigma^2_{-i}=(\sigma^{-2}_i-\tilde{\sigma}^{-2}_i)^{-1}$.

We then find, for each site, a new un-normalised Gaussian marginal with parameters
$\{\hat{\mu}_i,\hat{\sigma}_i,\hat{Z}_i\}$ which best approximates the product of the cavity distribution and the exact likelihood at each site.
\begin{equation}\label{eq: moment match}
\hat{q}(g_i):= \xi_i(g_i;\hat{\mu}_i,\hat{\sigma}_i,\hat{Z}_i)\approx q_{-i}(g_i)p(t_i|g_i).
\end{equation}
The parameters of the Gaussian $\hat{q}(g_i)$ are found by moment matching with the right hand side of equation (\ref{eq: moment match}). Finally, the site parameters $\{\tilde{\mu}_i,\tilde{\sigma}_i,\tilde{Z}_i\}$ of the likelihood approximation $\xi_i$ are obtained in turn from the updated moments of $\hat{q}(g_i)$. We refer the reader to \cite{Rasmussen2006} for a detailed derivation of the desired moments.

This procedure is carried out iteratively where in each sweep all the individual site functions are fitted and the sweeps are carried out until the convergence of the site parameters of $\xi_i$ for all the sites. In practice a fixed number of sweeps, say $20$, generally suffices for convergence. The converged site parameters are used to obtain the posterior latent function, at test locations \cite{Rasmussen2006}, 
\begin{equation}
p(\gb_*|\tb)\approx q(\gb_*|\tb)=\mathcal{N}(\bf{m_{g_*}},\bf{\Sigma_{g_*}}),
\end{equation}
where,
\begin{eqnarray}
\begin{aligned}
&\bf{m_{g_*}}=\mathbf{K_{*,g}}(\mathbf{K_{g,g}}+\tilde{\Sigma})^{-1}\tilde{\bv{\mu}},\\
&\bf{\Sigma_{g_*}}=\mathbf{K_{*,*}}-\mathbf{K_{*,g}}(\mathbf{K_{g,g}}+\tilde{\Sigma})^{-1}\mathbf{K_{g,*}}.
\end{aligned}
\end{eqnarray}
Substituting this value of $p(\gb_*|\tb)\approx q(\gb_*|\tb)$ into equation (\ref{eq: GP classifier class}), and approximating the sigmoidal function $sig(t_i| g_i)\approx\Phi(t_i |g_i)$ by a probit function\footnote{A probit function $\Phi(x)$ is the standard CDF of a normal distribution given by $\Phi(x)=\int_{-\infty}^{x} \mathcal{N}(y)dy$. EP gives us a very good Gaussian approximation as $\mathcal{N}(\bf{m_{g_*}},\bf{\Sigma_{g_*}})$.} we can evaluate the integral in equation (\ref{eq: GP classifier class}) to obtain the posterior class membership probability as
\begin{equation}
\begin{aligned}
\hat{\pi}:= p(\tb_*=+1|\tb)&=\int \Phi(\tb_*|\gb_*)\mathcal{N}(\bf{m_{g_*}},\bf{\Sigma_{g_*}})d\gb_*,\\
&=\Phi\left(  \frac{\bf{m_{g_*}}}{\sqrt{1+\bf{\Sigma_{g_*}}}}\right).
\end{aligned}
\end{equation}
We also use the site parameters to maximise $\ln(Z_{EP})$ to obtain the maximum likelihood estimates of the hyperparameters. 
From the likelihood approximations we directly obtain the marginal likelihood as the function of the site parameters given by
\begin{equation}\label{eq:classifier marginal}
\begin{aligned}
\ln (Z_{EP}) &\approx \ln \int \prod_{i=1}^{n} \xi_i(g_i;\tilde{\mu}_i,\tilde{\sigma}_i,\tilde{Z}_i)p(\gb|\phb)d\gb\\
&=\sum_{i=1}^{n}\ln \frac{\tilde{Z}_i}{\sqrt{2\pi}}-\frac{1}{2}\bv{V}^T\left( \mathbf{K_{g,g}}^{-1} +\mathbf{K_{g,g}}^{-1}\tilde{\Sigma}^{-1}\mathbf{K_{g,g}}^{-1}\right)
\bv{V}-\frac{1}{2}\ln\left|\mathbf{K_{g,g}}+\tilde{\Sigma}^{-1} \right|, 
\end{aligned}
\end{equation}
where,
\begin{equation}
\bv{V}=\left[ \mathbb{I} - \mathbf{K_{g,g}}\left(\mathbf{K_{g,g}}+\tilde{\Sigma}^{-1} \right)           \right]\mathbf{K_{g,g}} \tilde{\Sigma}\tilde{\bv{\mu}},
\end{equation}
and $\left|A \right|$ denotes the determinant of matrix $A$. 
Note that each iteration for maximizing $\ln Z_{EP}$ with respect to $\phb$ in equation (\ref{eq:classifier marginal}) requires the estimate of site parameters and thus a number of sweeps of EP. 
Thus the computational cost of maximising the marginal likelihood is much higher in this case compared to regression.

\subsection{Sparse approximations} \label{sec:GP FITC}
GP models suffer from high computational load for inference computations. For $n$ training points exact inference as used in GP regression requires $\mathrm{O}(n^3)$ effort while for EP approximation a sequence of $\mathrm{O}(n^3)$ operations are required.

There is an active line of research whose aim is to alleviate this computational bottleneck by using a sparse approximation of the true covariance function. 
Some of these methods are reviewed in \cite{quinonero2005unifying}. 
The common approach advocated by these methods is to use a set of $m$ inducing (or imaginary) inputs $\Xb_u$ with associated latent function $\bv{u}$ to reduce the computational load to $\mathrm{O}(nm^2)$. 
We denote the $n \times n$ covariance matrix between the training inputs as $\mathbf{K}$, the $m \times n$ covariance matrix between the inducing and training inputs as $\mathbf{K_{u}}$ and the $m \times m$ covariance matrix between the inducing inputs as $\mathbf{K_{u,u}}$. 
The most widely used approximation scheme \cite{quinonero2005unifying}, the FITC approximation $\hat{\mathbf{K}}$ to the full covariance $\mathbf{K}$ is given by
\begin{equation}
\mathbf{K} \approx \hat{\mathbf{K}} = \mathbf{Q}+\mathrm{diag}(\mathbf{K}-\mathbf{Q}),
\end{equation}
where $\mathrm{diag}(A)$ is a diagonal matrix whose elements match the diagonal of $A$ and the matrix $\mathbf{Q}$ is given by
\begin{align}
\mathbf{Q}&=\mathbf{K_{u}}^T\mathbf{Q_{u,u}}\mathbf{K_{u}},\\
\mathbf{Q_{u,u}}&=\mathbf{K_{u,u}}+\sigma^2_{n_{u}}\mathbb{I},
\end{align}
where $\sigma^2_{n_{u}}$ is the noise from inducing inputs. $\hat{\mathbf{K}}$ has the same diagonal elements as $\mathbf{K}$ and the off-diagonal elements are the same as for $\mathbf{Q}$. This sparse approximation was first introduced in \cite{snelson2006sparse} to scale the GP regression and later it was introduced in the context of a GP classifier in \cite{naish2008generalized} within the EP approximation.

\newpage
\section{Methods}
Having introduced the GP machinery we will now proceed towards applying the GP classification and regression to build a two-step emulator. 

\subsection{A GP classifier for segmenting the APD response surface: boundary detector} \label{sec:GP APD Classifier}

As mentioned in the previous section, our approach of using a classifier is primarily motivated by the idea of boundary detection and as a consequence segmentation of the input domain. 
However, unlike the previous domain segmentation attempts in the realm of computer experiments where a tree-based or non-stationary GP is constructed, in our application we can define a priori a set of possible labels to the APD\textsubscript{90} values (and the corresponding region of input space) based on the depolarisation/repolarisation pattern of the membrane voltage. 
The goal is to use a small number of simulated APD\textsubscript{90} values obtained by varying the parameters to train a classifier to predict the labels for a much larger set of test inputs with a quantifiable measure of uncertainty. 
We denote the inputs (here, scaling factors between zero and one applied to each maximal conductance of an ion current) as $\Rb=(R_1 , \ldots , R_D)$, where $D$ is the number of ion currents under consideration (and dimension of the input space) and each element of which can take any value between $[0,1]$. 
 
For the $n$-th input vector  our$\Rb_n$ our simulator $\mathcal{S}$ returns a set $\left\lbrace y_n,k_n\right\rbrace =\mathcal{S}(\Rb_n)$, where $y_n$ is the APD\textsubscript{90} value and $k_n\in \{1, 2, 3\}$ is a label associating the $n$-th input with any one of the three observed categories of action potential as shown in Figure~\ref{Figure:APbifurc}. 
We chose the following convention for labelling the action potential (see Fig. \ref{Figure:Various APs}): $k=1$ for no-depolarisation, $k=2$ for a valid action potential, and $k=3$ for no-repolarisation. 
Note that the simulator only returns an APD value in $y_n$ when the input is within the valid AP region $k=2$, and an error code denoting which of the other regions it is in otherwise.

As our problem is inherently a multi-class classification we adopt a One-versus-Rest (OVR) method of classification.
Using OVR we build one binary GP classifier, as introduced in section \ref{sec:GP Classification}, for each class $k\in \{1, 2, 3\}$ with associated labels $\tb^k \in \left\lbrace +1,-1\right\rbrace$ ($+1$ for the $k$-th class and $-1$ for the rest of the classes), to predict the probability 
\begin{equation}\label{eq:binary GP classifier}
\pi^{k}_*:= p(t^{k}_*=+1|R_*,\Rb,\tb^k),
\end{equation}
that a test input $\Rb_*$ given the training inputs $\Rb=\left(\Rb_1, \ldots, \Rb_n \right)$ and OVR labels $\tb^{k}$ belongs to the $k$-th class. The predicted class label $k_*$ for the test input $\Rb_*$ is then simply the most likely class:
\begin{equation}\label{class predict}
k_*=\underset{k}{\mathrm{argmax}}(\pi^{k}_*).
\end{equation}

\subsection{GP regression response surface prediction} \label{sec:GP emulator}
Again we consider a simulated dataset where $\left\lbrace \Rb_n,y_n\right\rbrace =\mathcal{S}(\Rb_n)$ is the $n$-th input-output pair which gives rise to a valid action potential and the associated APD\textsubscript{90} value returned by the simulator. 
We wish to learn a latent function $\fb$ that is an emulator of the simulator $\mathcal{S}$. 
Thus we have the output of the emulator $y_n$ (the APD\textsubscript{90} values) given by
\begin{equation}
y_n=\fb(\Rb_n).
\end{equation}
Now we place a zero mean GP prior on $\fb$ as
\begin{equation}
\fb \sim \mathcal{N}(0,K(\Rb,\Rb';\phb))
\end{equation}
where $K(\Rb,\Rb';\phb)$ is a covariance function parametrized by hyperparameters $\phb$. 
Note that this covariance is separate from the classifier covariance although we may use the same kernel function.  

Given the training data $\left\lbrace \Rb_*,\yb\right\rbrace$, the posterior mean at a new test point $f_* := f(\Rb_*)$ is given by (using equation (\ref{eq: GP post meanvar}))
\begin{equation}\label{GP emulator cond mean}
\mu(f_*)=K(\Rb,\Rb_*)- K(\Rb,\Rb)^{-1}\yb,
\end{equation}
and the posterior variance as (again using equation (\ref{eq: GP post meanvar}))
\begin{equation}\label{GP emulator cond var}
Var(f_*)=K(\Rb_*,\Rb_*)- K(\Rb,\Rb_*)^TK(\Rb,\Rb)^{-1}K(\Rb_*,\Rb).
\end{equation}

The hyperparameters are obtained as maximum likelihood estimates by maximizing the log marginal likelihood of the GP given by (following equation (\ref{eq:marginal likelihood}))
\begin{equation}
\begin{aligned}
\log P(\mathbf{y}|\phb)=-\frac{1}{2}\yb^{T}K(\Rb,\Rb)^{-1}\yb-\frac{1}{2}\log \left|K(\Rb,\Rb)\right|-\frac{1}{2}\log (2\pi).
\end{aligned}
\end{equation}

\subsection{Two-step emulator} \label{sec:Two step emulator}
We combine the boundary detector (using GP classification) and surface emulator (using GP regression) in a sequential manner to design a two-step emulator. 
In the training phase we use the simulator to create a training dataset of $n$ points:~$\mathcal{D}_{train} = \left\lbrace \left( \Rb_i, y_i, k_i\right) \mathrm{for}~ i=1, \ldots, n \right\rbrace$. 
We draw the values of $\Rb_i$ from  $\mathcal{U}(0,1)$. 
We then learn the GP hyperparameters associated with the boundary detector (with the OVR method using binary GP classifiers) and the surface emulator using the training dataset $\mathcal{D}_{train}$. 
Note that we use a subset of $\left\lbrace \Rb_i, y_i \right\rbrace $ for training the surface emulator. 
This subset contains only those inputs that generate a valid action potential --- that is, all training points in this subset have the same associated class label $k=2$. 

In the test/prediction phase for test input vector $\Rb_*$ we first use the boundary detector and obtain the class labels $k_*$ which we subsequently use to segment those test inputs into three domains: $\Rb^{NoDep}_*$ for which the membrane potential does not depolarise (that is no AP is generated); $\Rb^{AP}_*$ where we observe an AP; and $\Rb^{NoRep}_*$ where the membrane potential does not repolarise after the occurrence of an AP as shown in Fig.~\ref{Figure:Various APs}. 
Since we are interested in the AP region we pass $\Rb^{AP}_*$ to the surface emulator to obtain the posterior predictive given by (using equation (\ref{GP emulator cond mean}))
\begin{equation}\label{eq:APD pred}
y^{AP}\approx\hat{f}_*\sim \mathcal{N}\big(\mu(f_*),Var(f_*)\big).
\end{equation} 

\subsection{Choice of GP covariances} \label{sec:GP covariances}

In order to use both the classifier and surface GP one has to choose a suitable covariance function a-priori which embeds our prior assumption about the function that we are trying to model with a GP. 
For the classifier GP we use a squared exponential covariance given by equation (\ref{eq:squarred_exp}) and for the surface GP we use the rational quadratic covariance function given by
\begin{equation}
k_{RQ}(\Rb_i,\Rb_j)=\nu^2\left( 1+ \frac{\|\Rb_i-\Rb_j\|^2}{2\alpha l^2} \right)^{-\alpha},\label{eq:modified_squarred_exp}
\end{equation}
where $\{\nu,\alpha,l\}$ are the covariance hyperparameters. This covariance is equivalent to adding many squared exponential covariance with different lengthscales. 
As a result we expect to see functions varying across different lengthscales. 
Our prior intuition about the APD response surface is that it is smoother away from the boundary and varies much more rapidly near the boundary. 
Thus we have used this covariance function to accommodate different lenghtscales (degree of variation). 
Here the hyperparameter $\alpha$ determines the relative weighting of large-scale and small-scale variations.

We have used the GPML package \cite{rasmussen2010gaussian} called from MATLAB to implement the surface and boundary GPs. 
That code is available from \texttt{http://www.gaussianprocess.org/gpml/code/matlab/doc/}, and our simulator and emulator codes are available as described in Section~\ref{sec:materials}.

\subsection{Active learning} \label{sec:Active learning}
Since we are using a probabilistic framework for both classification and regression, we can exploit the uncertainty associated with the predictions to choose the training inputs using some form of adaptive scheme, as opposed to picking training points at random. 
This is known as \emph{active learning} \cite{krause2007nonmyopic}.
The main idea behind active learning is to sequentially add inputs to the training set to reduce uncertainty in predictions away from the training locations. 
Choosing the inputs actively has the potential to significantly reduce the required training budget, i.e.\ the number of simulations needed to generate $\mathcal{D}_{train}$. 
This is what we turn to next.

\subsubsection{Active learning for boundary detection using GP classification} \label{sec:Active learning for boundary detection}

\begin{figure}[!htb]
  \centering
  \includegraphics[width=\textwidth]{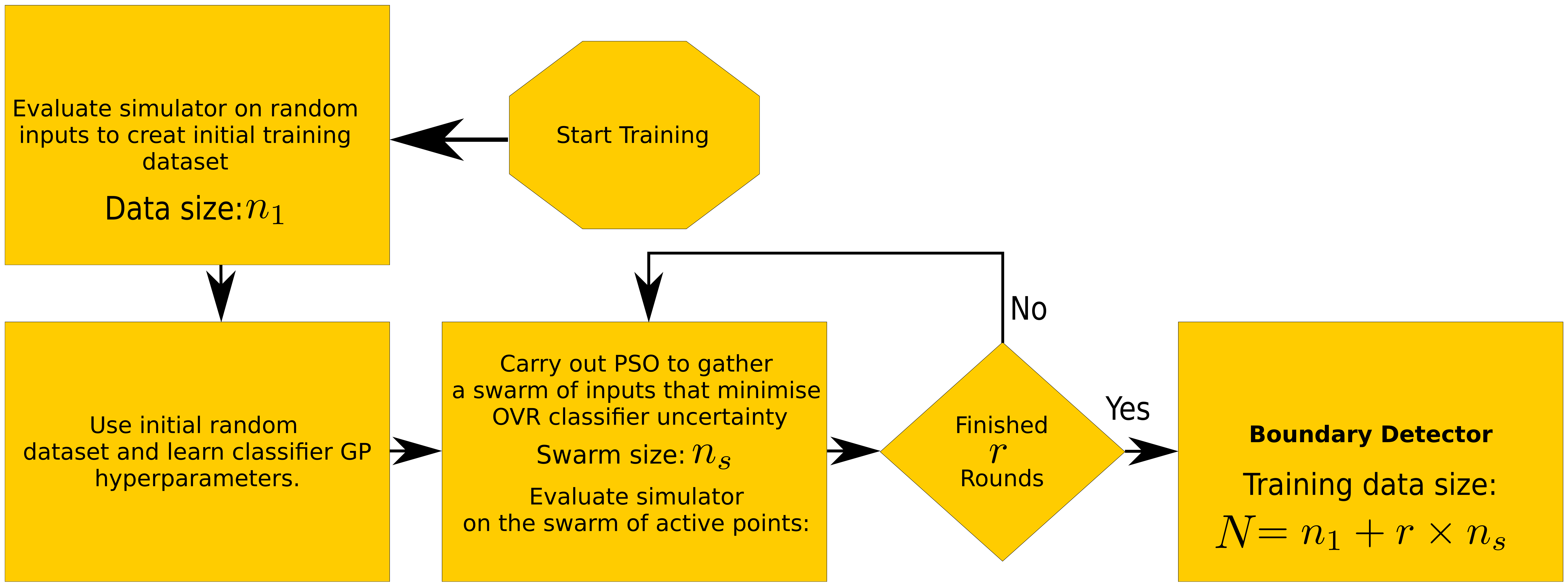}
  \caption{\textbf{Steps involved in active learning for the boundary detector}. 
  	The algorithmic settings, supplied by the user, for this active learning process are the initial data size $n_1$, number of rounds $r$ and the swarm size $n_s$. 
  }
  \label{Figure:activeClass}
\end{figure}

To carry out active learning of a GP classifier our goal is to augment a set of $n_1$ initial training inputs $\Rb_{\emptyset}$ with either a single new training point $\Rb_*$ or a set of  such points $\left\lbrace \Rb_{new_j}\right\rbrace_{j=1,\ldots,n_s}$ of size $n_s$ that convey more information about the boundaries in comparison to an equal number of randomly chosen points. 
We would like to point out that previously we have denoted a test point as $\Rb_*$ where the subscript $*$ denotes a point where we draw predictions from a GP. 
In this context, the same notation applies to those points where we draw predictions during the active learning process. 
We do this by finding those inputs about which the GP classifier is most uncertain, by using a suitable criterion to quantify this uncertainty. 
Information theoretic criteria such as the conditional entropy have been used to quantify uncertainty and carry out active learning for a GP classifier previously \cite{houlsby2012collaborative}. 
To quantify uncertainty through conditional entropy, we need to evaluate posterior expectations that do not have a closed form (due the the sigmoidal likelihood of the GP classifier). 
Gaussian approximation is known to work well for a binary GP classifier \cite{houlsby2012collaborative} and is considered a state-of-the-art technique to carry out active learning for a classifier GP. 
Although it is possible to extend the active learning approach proposed in Houlsby \emph{et al.}\ \cite{houlsby2012collaborative} to the multi-class OVR method, we propose an alternative quantification of uncertainty using the posterior prediction probability as:
\begin{equation}\label{certainty}
c(\Rb_*)={\mathrm{max}}(\pi^{k}_*)-{\mathrm{max}}(\pi^{k_{-}}_*),
\end{equation}
where $\pi^{k}_*$ is the classification probability for the most-likely class, given by equation (\ref{eq:binary GP classifier}), and the second term is the probability of classification into the second-most-likely class. 
We term $c$ the \emph{certainty} of classifier predictions as $c=1$ means the classifier is absolutely certain and $c=0$ represents equal probability of being in either of the two most likely classes. 
We are essentially characterising the regions in the input space that lead to an overlap of possible class distributions, and these are quantified as $c \rightarrow 0$.
We propose to use a particle based optimisation algorithm to find regions of minimum certainty, and (after this has converged) to add $n_s$ new points returned by the particle based optimiser to our training set to obtain the actively learnt input set $\Rb=\Rb_{\emptyset} \cup \Rb_{new}$. 
Moreover, we can carry out this procedure sequentially while using the optimiser at each step to find $\Rb_{new}$ and then setting $\Rb_{\emptyset}=\Rb_{\emptyset} \cup \Rb_{new}$. 
Repeating this for $r$ rounds we obtain the active set of inputs $\Rb$ consisting $n_3=r \times n_s$ new active training points, and thus a total training set at which the full simulator must be run of size $N=n_1+n_3$.
A flow diagram of active classifier learning is presented in Fig.~\ref{Figure:activeClass}. 
For carrying out the optimisation we use a particle swarm optimisation (PSO) algorithm \cite{eberhart1995particle,shi1999empirical} and to be synonymous with the terminology of PSO we will denote the set of points $\Rb_{new}$ considered at each round as a \emph{swarm} of possible training points.

Note that to generate distinct and useful training points for the classifier we purposefully refrain from running the PSO up to convergence, to maintain some distance (in input space) between each of the particles in the swarm. Thus in practice we stop the PSO iterations when the average certainty of the swarm goes below a threshold $\theta$. We found satisfactory performance by setting $\theta=0.5$.
To visualise the learning mechanism we evaluate this active learning scheme on a $2$-input simulator (and emulator) of the O'Hara model with inputs as: 
\begin{enumerate}[i)]
\item The sodium channel conductance scalings -- $R_{Na}$; and 
\item The hERG channel conductance scalings -- $R_{Kr}$; 
\end{enumerate}
with APD\textsubscript{90} as the output. 

\begin{figure}
  \centering
  \includegraphics[width=0.95\textwidth]{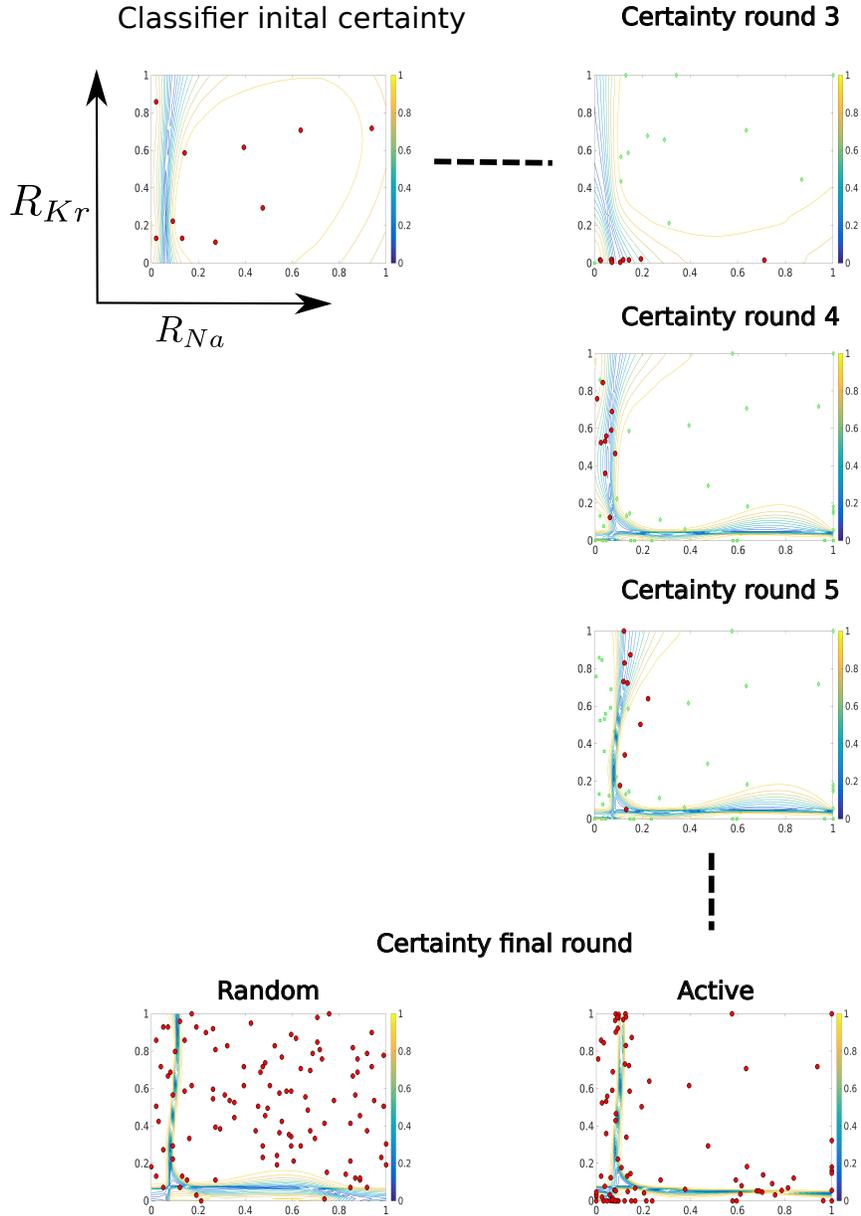}
  \caption{\textbf{Comparison of the measure of certainty $c$ in classification between actively and randomly adding training points}. 
  The certainty as defined by equation (\ref{certainty}) is shown as contour plots. The training points (accumulated thus far) are shown as red circles. 
  The darker shade of the contours represents the areas of least certainty spread along the boundaries (discontinuities) as estimated by the classifier. 
  We show three consecutive rounds of active learning $r=\{3,4,5\}$ in the right column. In the top left we show the initial certainty. In bottom row we compare the final certainty between active and random design.
  After round $r=4$ the active method starts discovering the boundaries and puts swarm points in this region. After the final round the active method chooses more points around the class boundaries, with most of the points placed in the region where the three classes meet. The region where three class boundaries intersect is the region of least certainty. 
  }
  \label{Figure:classCompare}
\end{figure}

We started with $n_1=10$ initial training points drawn randomly over the input space. We ran $r=10$ rounds of active learning with a swarm size of $n_s=10$ and show the consecutive uncertainty contours for rounds $r=\{3,4,5\}$ in Fig.~\ref{Figure:classCompare}. 
We also carried out a random design by adding $10$ points at random to the initial $10$ training points, and compare the certainty with that of active learning. 
The contour plots are shown in Fig.~\ref{Figure:classCompare}. 

It is evident from the contour plots for active learning that after round $r=4$ the classifier is able to get an estimate of the boundaries and the active points start to gather in the regions of low certainty. At the final ($r=10$) round the active learning method is able to increase the certainty significantly more than the random design.

\subsubsection{Active learning for surface emulation} \label{sec:Active learning for surface emulation}

Active learning for GP regression is a well studied topic and different schemes have been proposed.
The schemes differ mainly in the criteria with which uncertainty is quantified. The most widely used of these is the entropy criterion \cite{mackay1992information,seo2000gaussian,shewry1987maximum}. Alternative criteria such as mutual information \cite{krause2008near} and integrated variance \cite{gorodetsky2016mercer} have been proposed recently. 
Note that actively picking training points based on the entropy as well as mutual information is a $NP$-hard problem \cite{ko1995exact} and thus greedy algorithms are used while using any of these information theoretic criteria. 
Note that optimal algorithms have also been proposed \cite{krause2007nonmyopic,hoang2014nonmyopic} recently that integrate active learning with covariance hyperparameter learning, so that the selection of new training locations are optimal for carrying out Bayesian inference of the hyperparameters.

Most active learning methods have been used in spatial modelling where i) the training locations are 2D grids and ii) the number of training and test data points are much smaller than we can afford in our application where full simulator evaluations are relatively cheap (but not so cheap that an emulator is not desirable). 
In our application, the dimensionality of the inputs will be over two or three, and thus using a full grid is impractical. 
For these reasons we approach the active learning problem using a greedy algorithm, for computational tractability, that utilises the entropy criteria.

In order to set up the active learning consider a pool of \emph{candidate points} $\left\lbrace \Rb_{o_j}\right\rbrace_{j=1,\ldots,N_c} $ which are locations spanning the input domain, drawn randomly, from which we choose training points based on a chosen uncertainty criterion.
Our aim is then to choose a new training point $\Rb_*\in\Rb_{o}$ such that it gives us more information about the domain than choosing $\Rb_*$ randomly, resulting in greater prediction accuracy at lower simulation cost. 
This can be achieved by quantifying the uncertainty associated with the point $\Rb_*$, having observed a small initial dataset $\Rb_{\emptyset}$ consisting of $n_1$ points, where $n_1 \ll N_c$ for which the simulator returns a valid AP. 
Intuitively we want to choose $\Rb_*$ as points of maximum uncertainty. 
We can quantify this uncertainty using posterior conditional entropy between the latent function $f(\Rb_*)$ and the observations $\yb_{\emptyset}=\mathcal{S}(\Rb_{\emptyset})$ returned by the simulator.
This conditional entropy is given by \cite{krause2008near} 
\begin{equation}
H(f(\Rb_*)|\yb_{\emptyset})=-\int p(f(\Rb_*),\yb_{\emptyset})\log p(f(\Rb_*)|\yb_{\emptyset}) df(\Rb_*)d\yb_{\emptyset},
\end{equation}
which can be evaluated in closed form, since $p(f(\Rb_*)|\yb_{\emptyset})$ is the posterior Gaussian density of the surface GP, as
\begin{equation}\label{eq:diff entropy}
H(f(\Rb_*)|\yb_{\emptyset})=\frac{1}{2}\log\big( \mathrm{Var}(f(\Rb_*))\big)+\frac{1}{2}\log \big((2\pi)+1\big),
\end{equation}
where $Var(f(\Rb_*))$ is the posterior variance given by equation (\ref{GP emulator cond var}). 
We can now choose the point $\Rb_{*}$ from the $N_c$ candidate points $\Rb_{o}$ 
that has the largest conditional entropy:
\begin{equation}\label{eq: diff entropy}
\Rb_*=\underset{k}{\mathrm{argmax}}\big[ H\big(f(\Rb_{o}|\yb_{\emptyset})\big)  \big]  
\end{equation}
In order to implement such a scheme we have to be aware of the discontinuities. 
We start with a small random training set evaluated using the simulator at $\Rb_{\emptyset}$ for the surface (discarding non-AP points). 
We use a small $n_1$ to minimise the simulation cost. 
We also obtain an initial estimate of the hyperparameters using $n_1$. 
To remove the non-AP points from the candidates in $\Rb_{o}$ we can use the classifier. 
Inclusion of the classifier alleviates full simulations of all the candidate points.
Carrying out the above procedure we get the new training set as $\Rb=\Rb_{\emptyset} \cup \Rb_*$. We can also do this sequentially by setting $\Rb_{\emptyset}=\Rb_{\emptyset} \cup \Rb_*$ and then again finding one new point $\Rb_*$ using equation (\ref{eq: diff entropy}). 

\begin{figure}[!htb]
  \centering
  \includegraphics[width=\textwidth]{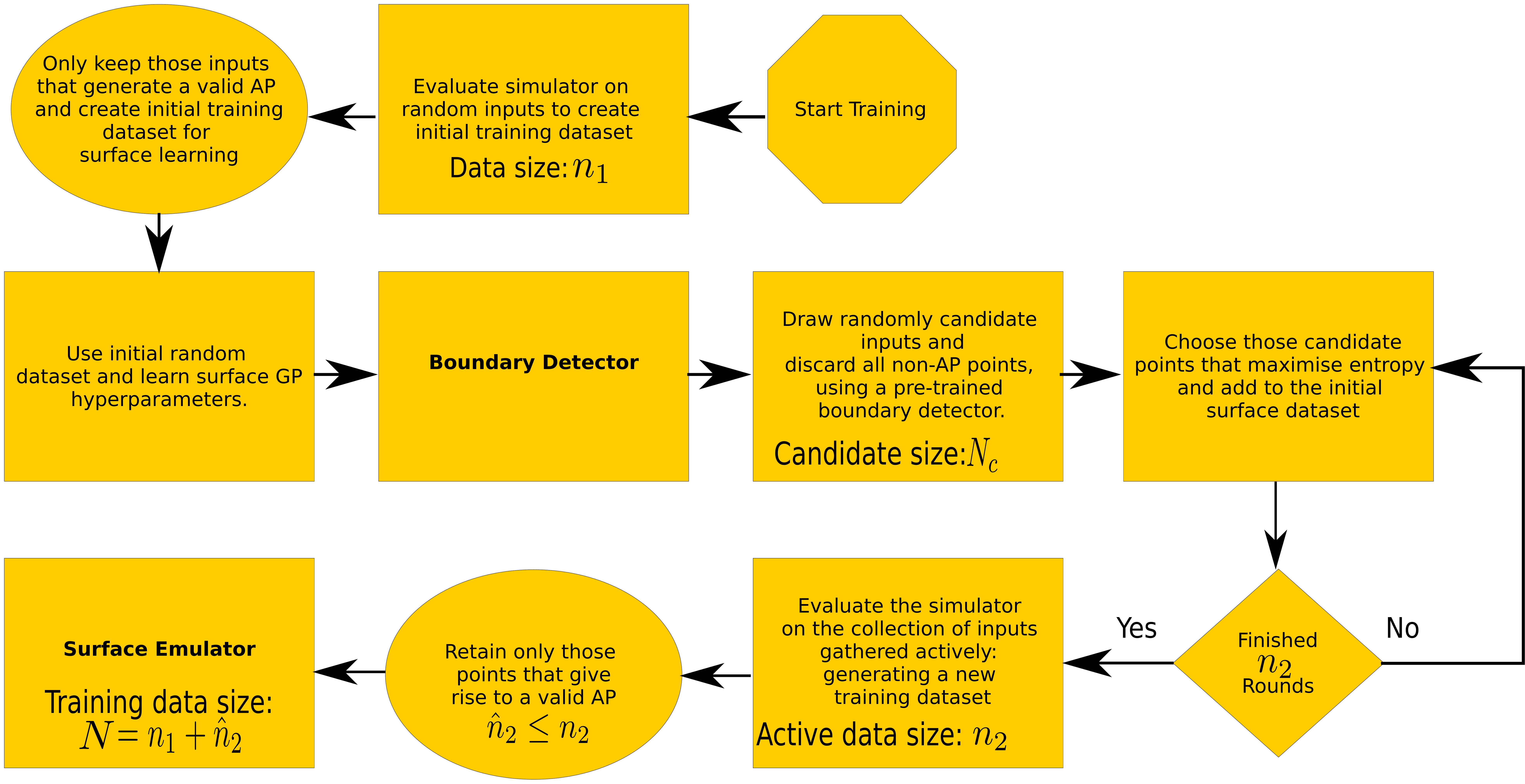}
  \caption{\textbf{Steps involved in active learning for the surface predictor}. 
  We have used the oval shapes to indicate steps where we are removing non-AP points from the training set. 
  The rest of the shapes used here follow standard flowchart notations. 
The algorithmic settings that needs to be supplied by the user are the initial data size $n_1$, the candidate set size $N_c$ and the number of sequential rounds $n_2$.
}
  \label{Figure:activeSurf}
\end{figure}

We can carry on this iterative procedure for $n_2$ rounds to collect $n_2$ active training input points. Note that due to misclassification some of the candidate points in $\Rb_{\emptyset}$ may be in non-AP regions.
When we evaluate the simulator on the $n_2$ active points we remove these non-AP points before forming the surface GP training set. 
Thus the resulting set of active points is of size $\hat{n}_2 \leq n_2$ and our new training set after carrying out active learning consists of $N=n_1 + \hat{n}_2$ training points. Also note that the classifier is used only once on the entire candidate set before adding an active point.
The various steps involved in the surface active learning are summarised and presented in Fig.~\ref{Figure:activeSurf} as a flow diagram.

\begin{figure}[!htb]
  \centering
  \includegraphics[width=0.85\textwidth]{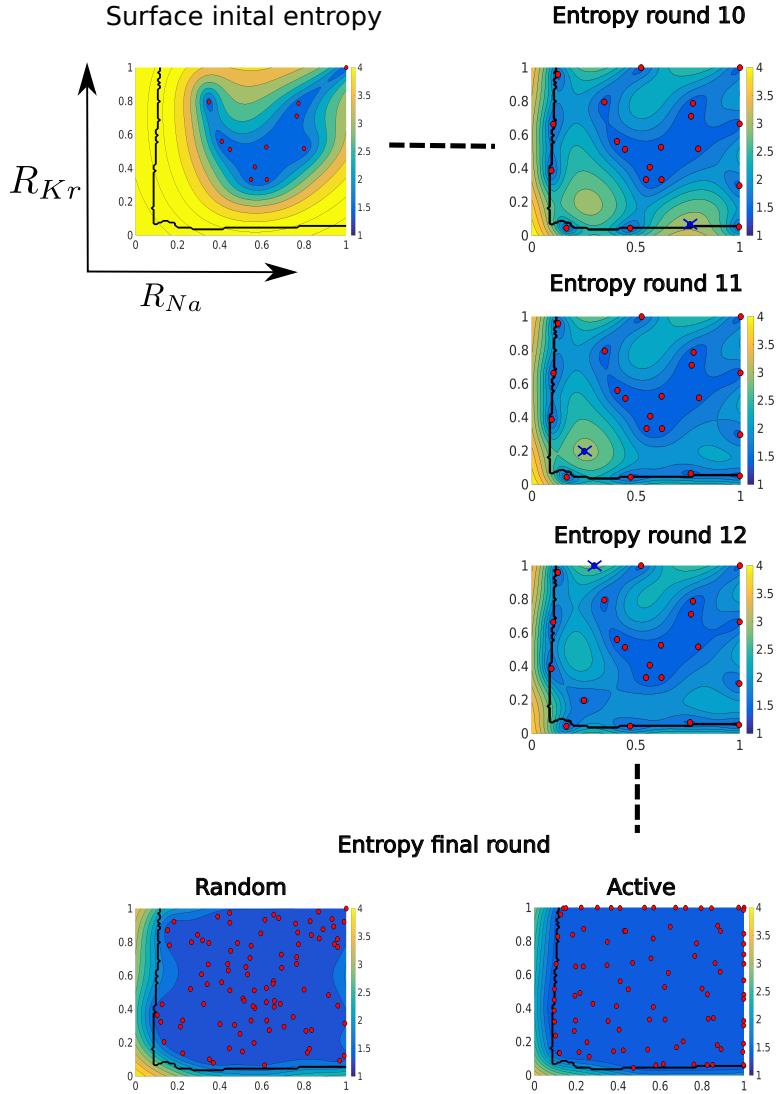}
  \caption{\textbf{Comparisons of entropies from active vs.\ random training of the surface GP}. 
  The entropy defined by equation (\ref{eq: diff entropy}) is shown as a contour plot. 
  A darker shade represents less entropy and thus less uncertainty in surface predictions. 
  The black line demarcates the real boundary (obtained using a $100 \times 100$ grid for visualisation, and shown in Fig.~\ref{Figure:APDsurface}) between the non-AP and valid AP regions. 
  The red circles show all the training points accumulated at specific stages of training in both active and random learning schemes. 
  The active points picked during intermediate rounds are shown as blue circles with black cross-hair.
  We show three consecutive rounds of active learning $r=\{10,11,12\}$ in the right column. 
The next training point is always placed in the most uncertain (high entropy) region on the input space. 
In the top left we show the initial uncertainty on $n_1=11$ training points. 
In the bottom row we compare the final uncertainty between the active and random schemes after $n_2=90$ rounds. 
  The active method is able to reduce the uncertainty noticeably over the input domain after the final round. Also note that in the intermediate rounds such as $r=\{10,11,12\}$ some active points are picked in the non-AP region. This happens due to misclassification of the candidate points. 
  However, after we finish the active learning these points are discarded based on the actual simulator evaluation.  
  }
  \label{Figure:surfCompare}
\end{figure}

We again set up the $2$ input problem exactly as illustrated previously for classification. 
We started with $n_1=11$ initial training points drawn randomly over the input space and carried out active learning for $n_2=90$ rounds using candidate inputs $\left\lbrace \Rb_{o_j}\right\rbrace_{j=1,\ldots,N_c}$ of size $N_c=10{,}000$. 
We also carried out a random design by adding points sequentially to the training set starting with the initial $n_1$ points for a total of $90$ rounds. 
Fig~\ref{Figure:surfCompare} shows a comparison of consecutive entropies during active learning rounds $\{10,11,12\}$. 
We also show in the same figure (bottom row) the comparison of entropies between active learning and random design at the final round. 
We observe during rounds $\{10,11,12\}$ that the active points (blue circles with cross-hair in Fig~\ref{Figure:surfCompare}) are placed at places of high entropy. 
Active learning is able to reduce the entropies better than random design.

In our active learning scheme we restrict the candidate points to be within the valid AP region by i) before simulation filtering out non-AP candidate points using the classifier; and ii) after simulation, discarding any non-AP points that were misclassified. 
During the active learning process the entropies are higher in the regions near the boundaries as only a few candidate points (due to misclassification) exist in the non-AP regions. 
A well-studied pathology of entropy based active learning \cite{krause2007nonmyopic} is that a lot of training points are selected near the edges of the surface, that is the regions of highest entropies. 
As the active learning progresses the boundary regions emerge as regions of high uncertainty for the reasons described above.
This is something that we notice in the contour plots. 
However, this behaviour actually works in our favour for this application as the surface changes most rapidly near the boundary region and thus having training points in those regions leads to a better surface prediction. 

\FloatBarrier

\section{Results}\label{sec:Experiments}

We are going to test different aspects of the two-step emulator applied to predicting the APD response surface as obtained by simulating the O'Hara \cite{o2011simulation} cardiac electrophysiology model. 
We evaluate the performance of the emulator by observing the rate of decrease in prediction error, a \emph{learning curve}, of the surface and boundary GP respectively as we grow the size of the training dataset. 
We have designed two sets of tests.
In the first set, we evaluate the learning curve of the surface and classifier GP without active learning, which we denote as random learning. 
In the second, we evaluate similar learning curves using the active learning schemes described in sections \ref{sec:Active learning for surface emulation} \& \ref{sec:Active learning for boundary detection}.

Note that the error in the surface GP cannot be evaluated at points that belong to either of the non-repolarising or non-depolarising regions.
To circumvent this, we only use test points that are associated with a valid AP to obtain learning curves (in both random and active learning experiments) for the surface GP, but we also track the misclassification rate of all points.

We also compared the learning curves in the first set of experiments, the random case, with the learning curves of a look-up table based interpolator used within \cite{mirams2014prediction} and subsequently in a web-based APD prediction application \cite{williams2015web}. 
This interpolator uses a look up table and a space partitioning algorithm to interpolate a range of biomarkers including the APD. 
Following the emulator tests we only test the interpolator on inputs generating a valid AP while comparing the surface predictions. 
To test the classifier GP's performance we use inputs on the entire domain.

\subsection{Simulator setup} \label{sec:Simulator setup}
As mentioned previously, our simulator is the ventricular action potential model from \cite{o2011simulation}. 
For the experiments below, the 4D input to the simulator is the conductance scalings of four ion currents: 
\begin{enumerate}[i)]
\item Fast sodium channel conductance scaling --- $R_{Na}$; 
\item Rapid delayed rectifying potassium channel conductance scaling --- $R_{Kr}$;
\item Slow delayed rectifying potassium channel conductance scaling --- $R_{Ks}$;
\item L-type calcium channel conductance scaling --- $R_{CaL}$. 
\end{enumerate}
The output of the simulator is the APD\textsubscript{90} value under these channel scalings. 
For each evaluation of the simulator we pace the cardiac model with $100$ 1\,Hz paces (using the stimulus defined in the CellML file/original \cite{o2011simulation} paper) in order to allow the state variables to settle towards their 1\,Hz limit-cycle (a larger number may be required in practice). 
Thus for each input combination the simulator calls the underlying ODE solver $100$ times. 
We have used the Chaste cardiac modelling package \cite{mirams2013chaste} to implement the simulator using a CellML representation \cite{Cooper2015} of the model. 
Code is openly available as described in Section~\ref{sec:materials}.

\subsection{Surface and classifier GP learning curves: random method} \label{sec:Surface and boundary learning curves: random method}
For this experiment we used the simulator (setup as described previously) to generate a training, $\mathcal{D}_{train}$, and a test, $\mathcal{D}_{test}$, dataset each containing a different $N=100{,}000$ points. 
For evaluating the surface GP we keep only those points, in both the training and test datasets, that are associated with a valid AP. 
To obtain the learning curve we generate a random subsample $\mathcal{D}^{*}{train}=\{500,\ldots,50{\,}000\}$ of the training data to fit the hyperparameters by MLE for both the surface and boundary GPs and draw prediction on the entire 100,000 test points. 
Denoting the test dataset as $\mathcal{D}_{test}= \left\lbrace \left( \Rb_i, y_i, k_i\right) , i=1, \ldots, N \right\rbrace$ we define the surface prediction error $\mathrm{E}_{surface}$  as the mean absolute error in the APD\textsubscript{90} given by 
\begin{equation}\label{surface error}
\mathrm{E}_{surface}=\frac{1}{N_{\mathrm{AP}}}\sum_{i=1}^{N_{\mathrm{AP}}} |y_i-\mu(f_i)|,
\end{equation}
where $N_{\mathrm{AP}}$ is the number of test points associated with a valid AP, $\mu(f_i)$ defines the posterior mean prediction of the surface emulator GP, and $|.|$ defines the absolute value. 
The error for the boundary detector $\mathrm{E}_{boundary}$ is defined as the percentage misclassification rate given by
\begin{equation}\label{boundary error}
\mathrm{E}_{boundary}=\frac{100}{N}\sum_{i=1}^N\mathbb{I}(k_i\neq \widehat{k_{i}}),
\end{equation}
where $\widehat{k_{i}}$ is the most likely class prediction from the OVR classifier and $\mathbb{I}(\cdot)$ denotes the indicator function. 

Also note that we start using sparse covariances, using the FITC approximation, when number of training points is greater than 10,000 and 5000 for the surface and classifier GPs respectively. 
We used the FITC approximation with 1000 and 300 training points (see section  \ref{sec:GP FITC}) for the surface and classifier GPs respectively. 
This number of training points is sufficient to produce an approximation comparable (and superior) to that of the Look-Up-Table interpolator.

\begin{figure*}[!htb]
  \centering
  \subfigure[]{
    \includegraphics[width=2.5in,height=1.77in]{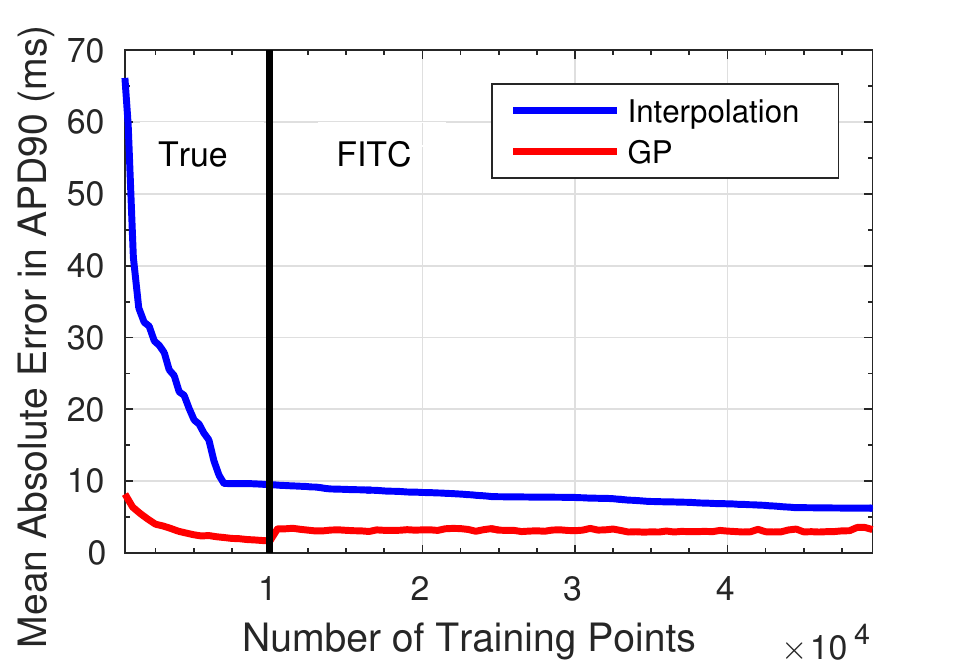}
    \label{Figure:surfErrorInterpVsGp}
  }  
	\subfigure[]{
    \includegraphics[width=2.5in,height=1.77in]{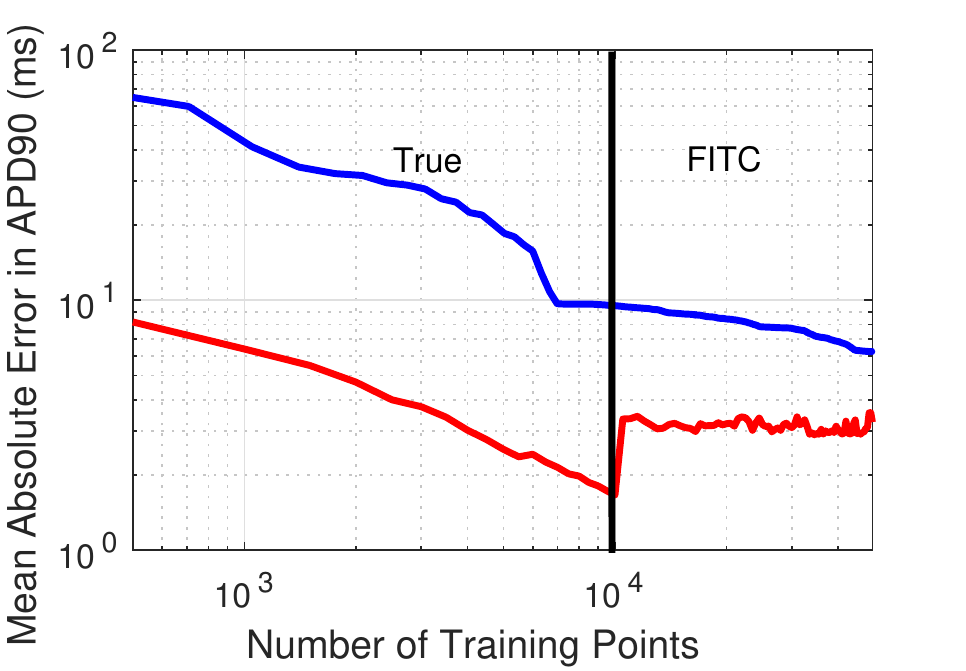}
    \label{Figure:surfErrorInterpVsGplog}
  } 
  \caption{\textbf{Learning curves for the surface GP}. a) Original axes, b) Logarithmic axes. 
  The black line distinguishes between the part of learning curves generated using the true covariances and the FITC approximations.  
  }
  \label{Figure:GPvsIntlsurfError}  
\end{figure*}

\begin{figure*}[!htb]
  \centering
  \subfigure[]{
    \includegraphics[width=2.5in,height=1.77in]{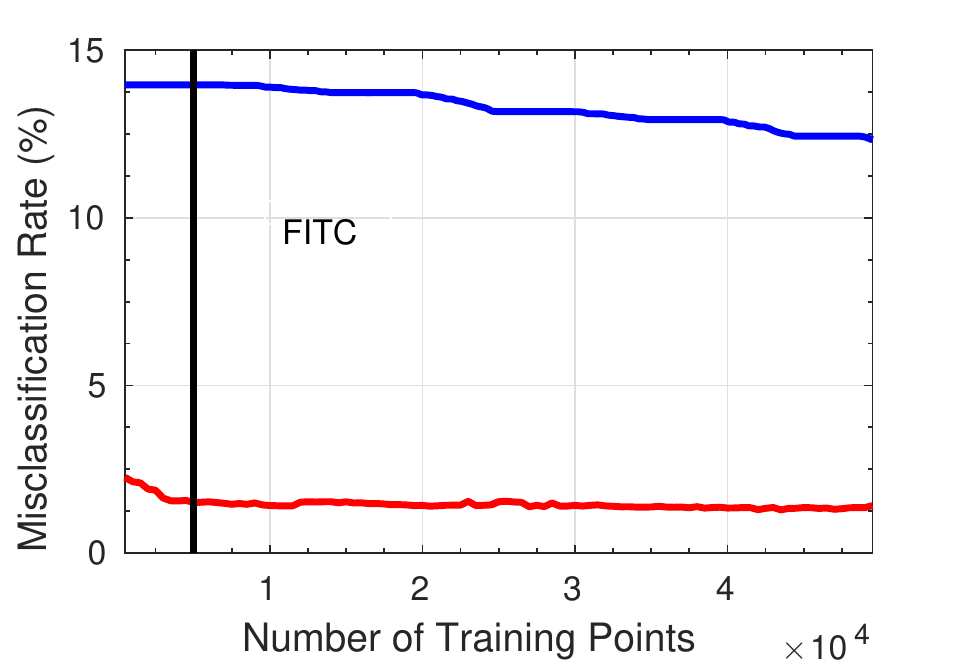}
    \label{Figure:classErrorInterpVsGp}
  }  
	\subfigure[]{
    \includegraphics[width=2.5in,height=1.77in]{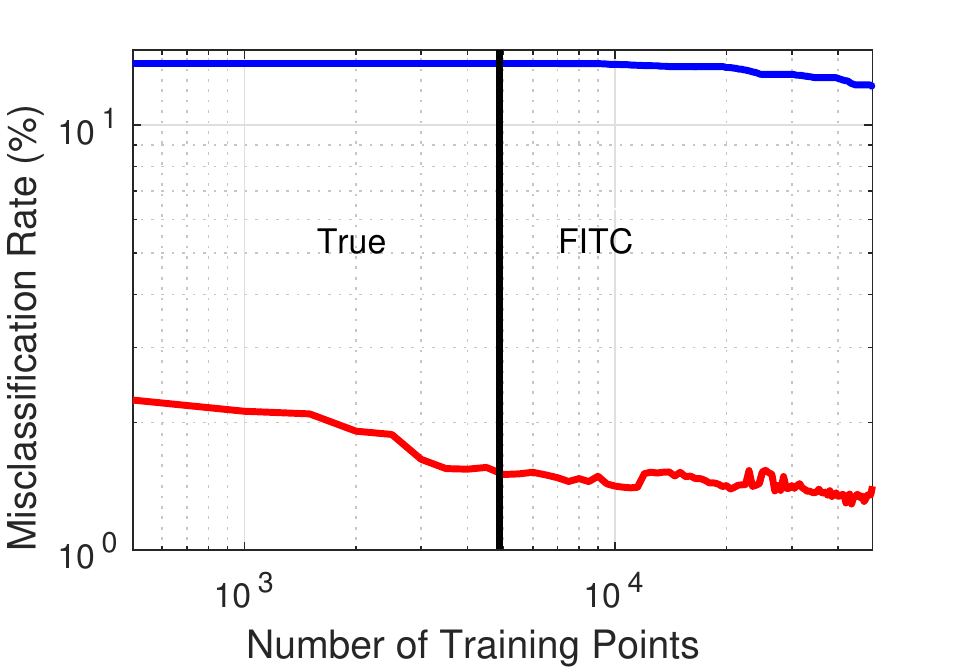}
    \label{Figure:classErrorInterpVsGplog}
  } 
  \caption{\textbf{Learning curves for the classifier GP}. a) Original axes, b) Logarithmic axes.
  	  The black line distinguishes between the part of learning curves generated using the true covariances and the FITC approximations.  
}
  \label{Figure:GPvsIntlclassError}  
\end{figure*}

We plot the learning curves for the surface and classifier GPs in Figs.~\ref{Figure:GPvsIntlsurfError} \& \ref{Figure:GPvsIntlclassError}. 
In both of these figures the black line distinguishes between the part of learning curves generated using the true and the FITC covariance.

Unlike the classifier GP learning curve (Fig.~\ref{Figure:classErrorInterpVsGplog}) the surface GP error stops decreasing as the FITC approximation is introduced and this error remains relatively constant despite the increase in training data size (see Fig.~\ref{Figure:surfErrorInterpVsGplog}). 
The error generated by the surface GP with a training data size of $10{,}000$ is the minimum error achievable, on the test dataset that we have used, and thus introducing more training points is futile in decreasing the error. 
In fact with the FITC covariance the error goes up due the introduction of the covariance approximation. 
The surface GP clearly outperforms the linear interpolator; this is expected as the interpolator employs a very simple function estimation compared to GP regression. 

From Fig.~\ref{Figure:classErrorInterpVsGp} it is evident that the classifier error decreases steadily at a faster rate up to $4000$ training points.
Although the error keeps on decreasing beyond this training data size, the rate of decrease is reduced noticeably. 
Also notice in Fig.~\ref{Figure:classErrorInterpVsGplog} that using a sparse covariance approximation does not appear to hinder the accuracy of the GP classifier. 

\subsection{Performance with active learning: surface emulation} \label{sec:Performance_with_active_learning:surface}

In section \ref{sec:Active learning for surface emulation} we gave an overview of the surface active learning scheme. 
Although entropy based active learning is a well studied method, doing so on a discontinuous surface brings about a new set of challenges so as to confine the learning scheme within the boundaries of discontinuities. 
This is achieved using the classifier. 
We start with an initial dataset $\mathcal{D}_{\emptyset}\in \mathcal{D}_{train}$ of size $n_1=500$, a random subsample of $\mathcal{D}_{train}$. 
We also fix the candidate set $\left\lbrace \Rb_{o_j}\right\rbrace_{j=1,\ldots,N_c}$ size as $N_c=10{,}000$. 
For evaluation purposes we train (learn the hyperparameters) of the surface and classifier GPs using $\mathcal{D}_{\emptyset}$. 
Note that for actual applications we will have an (actively) pre-trained classifier GP. We test this approach below in section \ref{sec:Performance of the two-step method}.
We also fix the number of active points to $n_2=2500$. 
With no misclassification the final training set size would be $N=n_1 + n_2=3000$. 
However, we end up with a training size $N_{\mathrm{AP}} < N$ after discarding the inputs with invalid APs. 

\begin{figure*}[!htb]
  \centering
  \subfigure[]{
    \includegraphics[width=5in,height=3in]{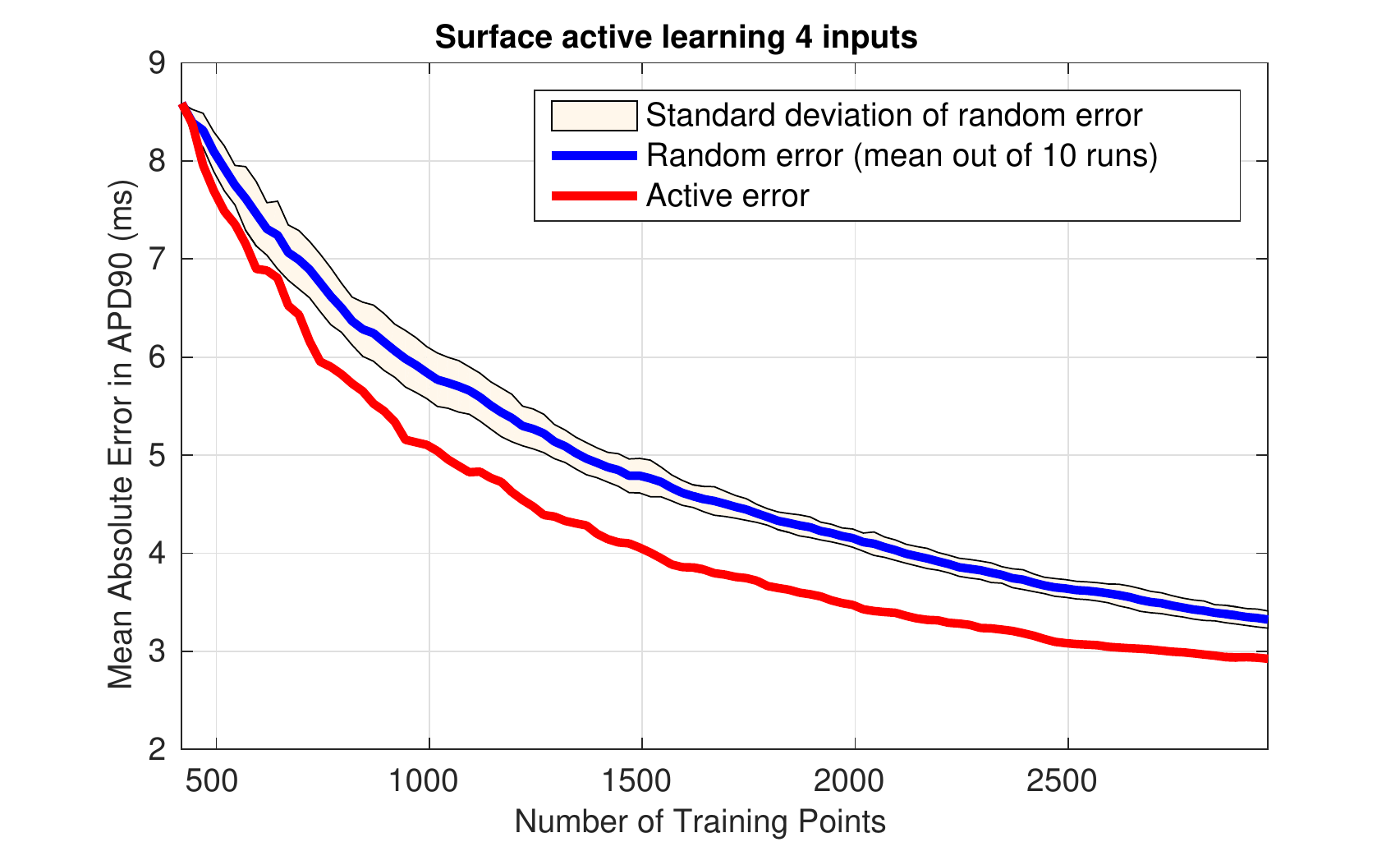}
    \label{Figure:surfaceAL4D}
  }  
  \subfigure[]{
    \includegraphics[width=5in,height=3in]{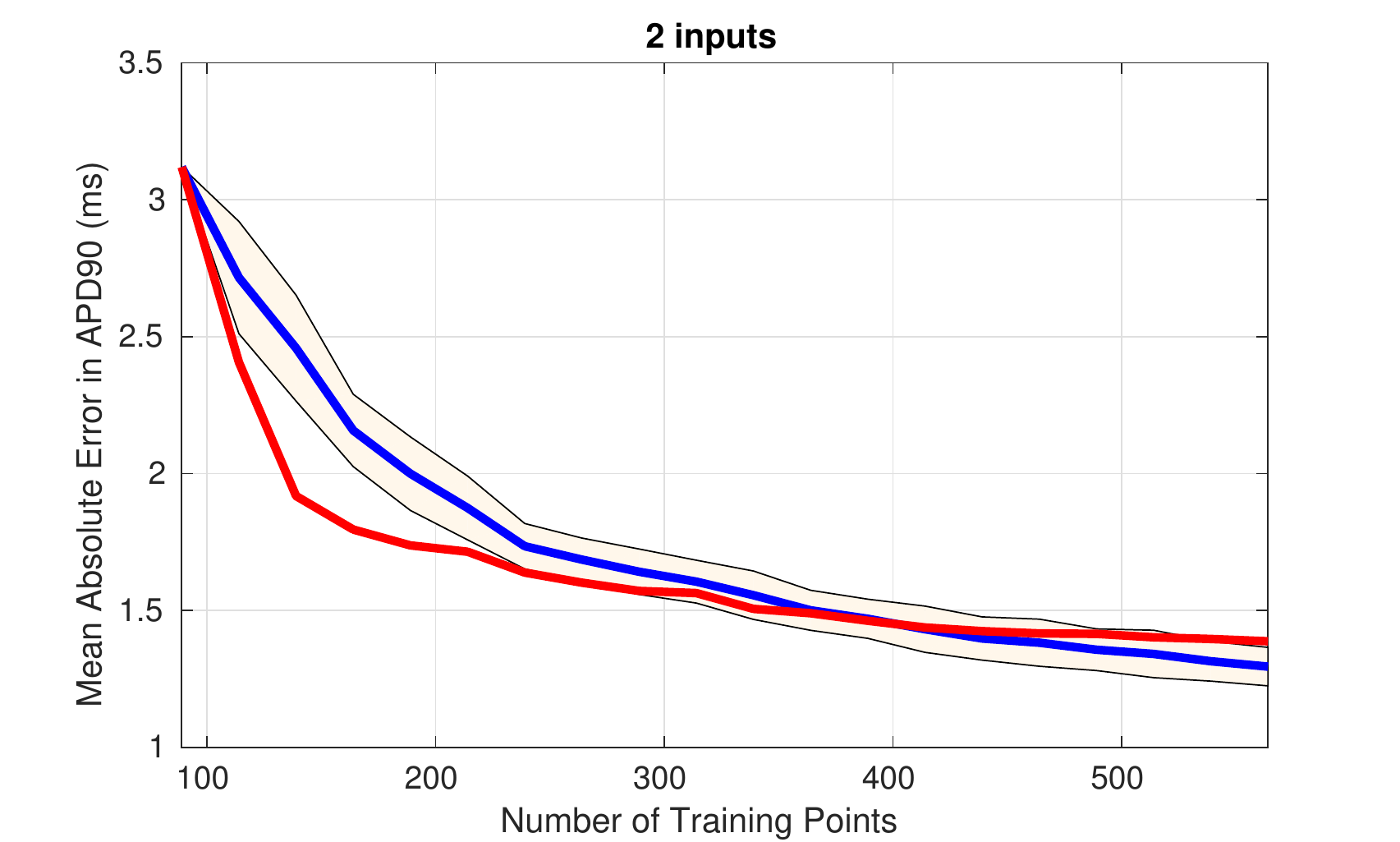}
    \label{Figure:surfaceAL2D}
  } 
  \caption{\textbf{Learning curves for surface active vs.\ random learning for the a) $4$-input $\&$ b) $2$-input problem}. The red line shows the error for training the GP using actively learnt inputs. The blue line shows the average, out of $10$ repetitions, GP error for randomly drawing training inputs. The shaded area shows the standard deviation of this error, reflecting the variability in performance if a random design is carried out.
  }
  \label{Figure:4DAl}  
\end{figure*}

To generate a learning curve we calculate the surface error using equation (\ref{surface error}) by sequentially generating predictions with the active dataset after the inclusion of $25$ new training points. Note alternatively we can evaluate the learning curve after inclusion of every other active point. However, to keep parity with the classifier learning curves (presented in section \ref{sec:Performance_with_active_learning:classifier}), evaluated on each new swarm of active points, we adopt the aforementioned frequency of evaluation.
The prediction is made on the test dataset $\mathcal{D}_{test}$, containing $N_{\mathrm{AP}}$ test points. 
To compare this with randomly adding training points to the initial set $\mathcal{D}_{\emptyset}$ we calculate the same sequential prediction errors by drawing $25$ new points randomly from $\mathcal{D}_{train}$, having only valid AP inputs, as opposed to actively learning them. We keep on adding these $25$ random points for $n_2/25=100$ rounds. 
To highlight the variability in randomly growing the training dataset we repeat the random learning exercise $10$ times.
Furthermore, we draw the predictions in both schemes with the set of hyperparameters learnt using the same initial dataset $\mathcal{D}_{\emptyset}$. 
We plot the resulting learning curves in Fig.~\ref{Figure:surfaceAL4D}. 
From Fig.~\ref{Figure:surfaceAL4D} it is evident that the active learning scheme outperforms the random update method. 
In Fig.~\ref{Figure:surfaceAL2D} we plot the learning curves for the $2$-input problem that we used for visualisations of the surface entropy in section \ref{sec:Active learning for surface emulation}. 
Only in this case we change the initial points size to $n_1=100$ to improve the hyperparameter estimate, and to compare with random design we keep adding $25$ new randomly drawn points sequentially to the initial training set as done above for the $4$-input problem.
We generated a test dataset containing $10{,}000$ points placed on a $2$-dimensional grid for this $2$-input case. 
We extended the rounds to $500$ in order to have a smoother learning curve. 
The swarm size for evaluating error is maintained as $25$.
\begin{figure*}[!htb]
  \centering
  	\subfigure[]{
      \includegraphics[width=5in,height=3in]{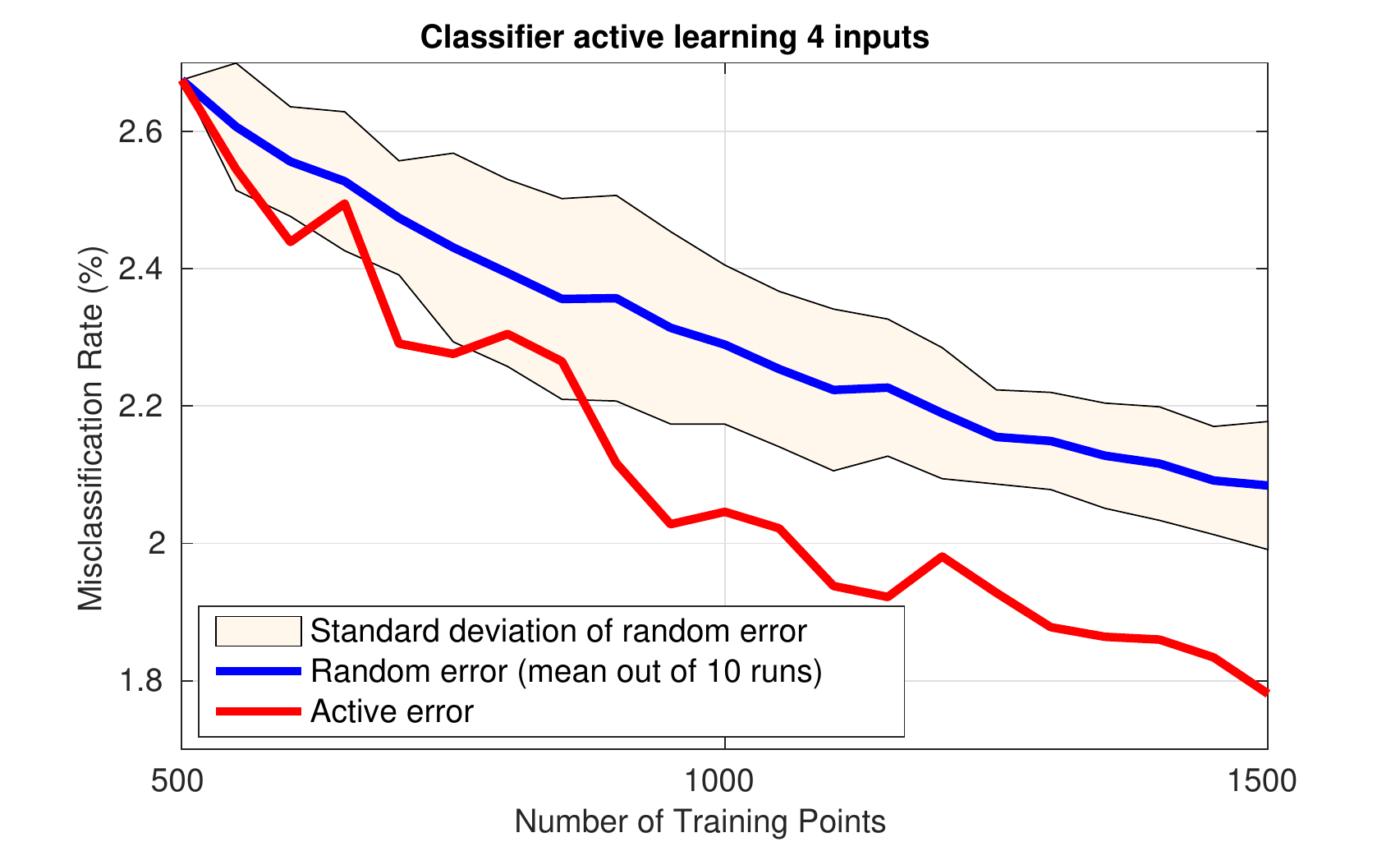}
      \label{Figure:classifierAL4D}
    }  
	\subfigure[]{
    \includegraphics[width=5in,height=3in]{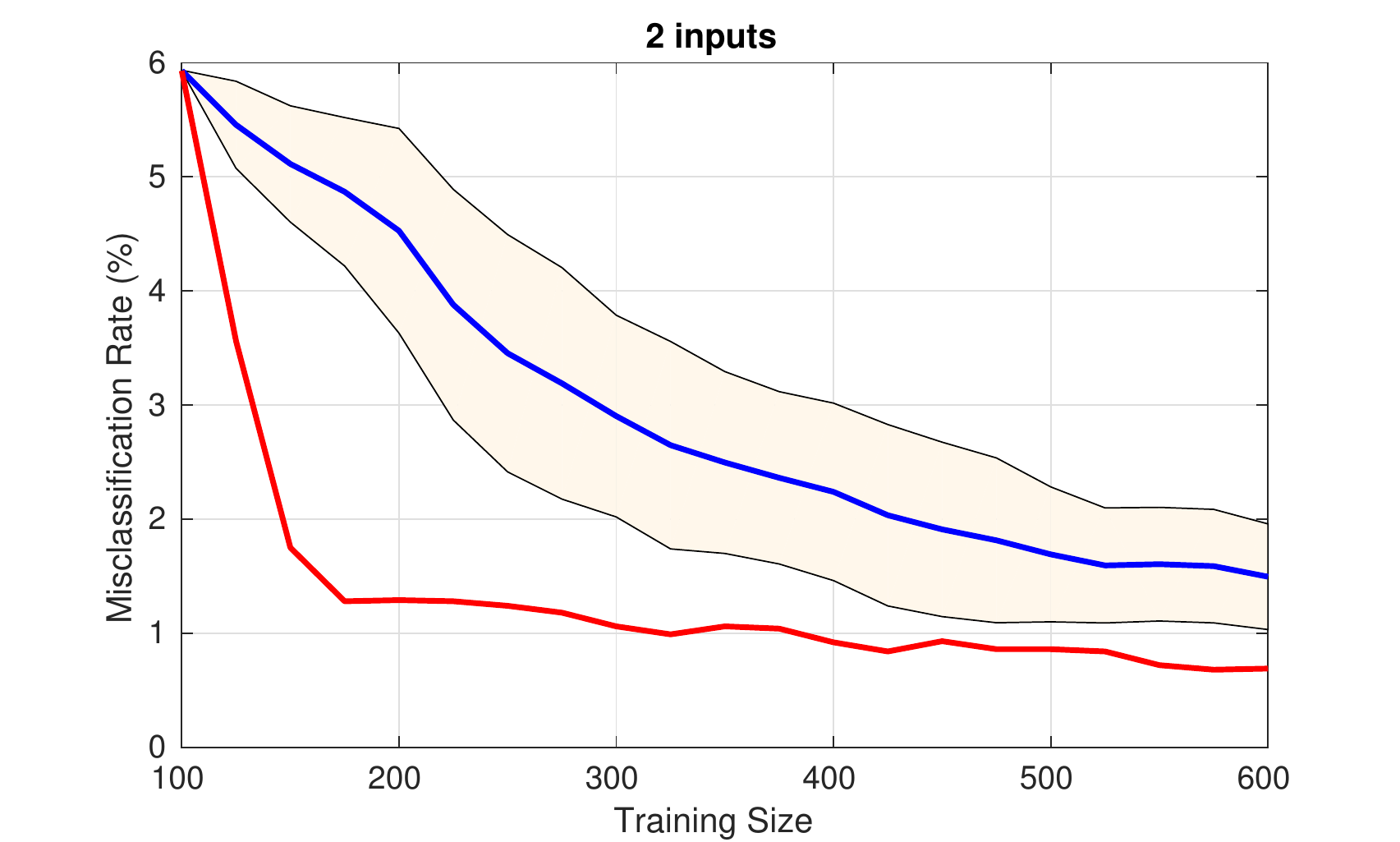}
    \label{Figure:classifierAL2D}
  } 
  \caption{\textbf{Learning curves for classifier active vs.\ random learning for the a) $4$-input $\&$ b) $2$-input problem}.}
  \label{Figure:2DAl}  
\end{figure*}
The increased accuracy in the 2-input case, for both active and random learning, results from the reduced dimensionality of the discontinuity surface. 
Also note that the error is reduced much faster during the earlier rounds of active learning than the later ones. 

The entropy criteria will eventually start picking all those points from the candidate set that are located near the surface boundaries (regions of higher entropy including the class boundaries), once a certain number of points are picked in the valid AP region to reduce the entropy significantly. 
Although it is difficult to quantify the number of points (in the valid AP region) required to reduce entropy optimally we start to see this effect for the simpler $2$-input problem in Fig.~\ref{Figure:surfaceAL2D}. 
Once the number of active training points is greater than $400$ the learning curve improvement slows. 
At this point the active learning scheme is adding points mostly around the surface boundaries, whereas for the random addition of training points more points are accumulated in the valid AP region and thus the random error keeps on decreasing. 
It is worth noting that in applications such as ours some of the boundary regions (no ion channel block in certain dimensions) are of particular interest, and extra accuracy there is beneficial.
Due to the nature of the entropy criteria we expect the same thing to happen in the $4$-input case but for a higher number of training points.

\subsection{Evaluating classifier active learning} \label{sec:Performance_with_active_learning:classifier}

We adopt the same procedure for testing the classifier active learning as for the surface. 
That is, we try to compare the misclassification error produced by actively growing the training dataset to that of a random design. 

We start with an initial dataset $\mathcal{D}_{\emptyset}\in \mathcal{D}_{train}$ of size $n_1=500$ and sequentially grow the training dataset using PSO as described in section \ref{sec:Active learning for boundary detection} to collect a swarm of inputs of size $n_s=50$ in each round.
We retain the same test dataset and we repeat this process for a 2-input problem set up exactly as in the surface active learning experiment. 
The random learning is carried out as previously, using $10$ randomly drawn swarms in each round from $\mathcal{D}_{train}$. 
For the $4$-input problem we perform $r=20$ rounds of PSO, collecting $n_3=20 \times 50=1000$ active input points, and thus $N=n_1 + n_3=1500$ training points in total. 
For the $2$-input case we start with $n_1=100$ initial points and restrict the swarm size to $25$ points, thus collecting $n_3=25 \times 20=500$ active input points and $N=n_1 + n_3=600$ training points in total. 
We retain the same test dataset which was used for evaluating the surface active learning in the $2$-input case. We use a cut-off value $\theta=0.5$ specified by the average certainty of the swarm particles to stop the PSO iterations. 

We plot the learning curves for the boundary classifier in Fig.~\ref{Figure:classifierAL4D}. 
The learning curve for the $2$-input problem is shown in Fig.~\ref{Figure:classifierAL2D}. 
We have created an animation (see Supplementary Material) visualising the first 8 rounds of PSO and subsequent changes to contour plots of the certainty.
In both these plots we see that the active error decreases much more rapidly than in the surface case. 
Furthermore, we notice that the variability (among the $10$ repetitions) of random errors for the classifier GP is higher than the surface error plots.

However, for the $2$-input case we see the same flattening of the active learning curve as in the surface active learning. This is because after going through $20$ rounds the active learning scheme manages to put the necessary amount of input points covering all the uncertain regions on the input space. 
Adding further inputs does not affect the overall uncertainty noticeably. 

\subsection{Learning times and swarm sizes} \label{sec:Performance with active learning}

\begin{figure*}[!htb]
  \centering
  \subfigure[]{
    \includegraphics[width=5in,height=3in]{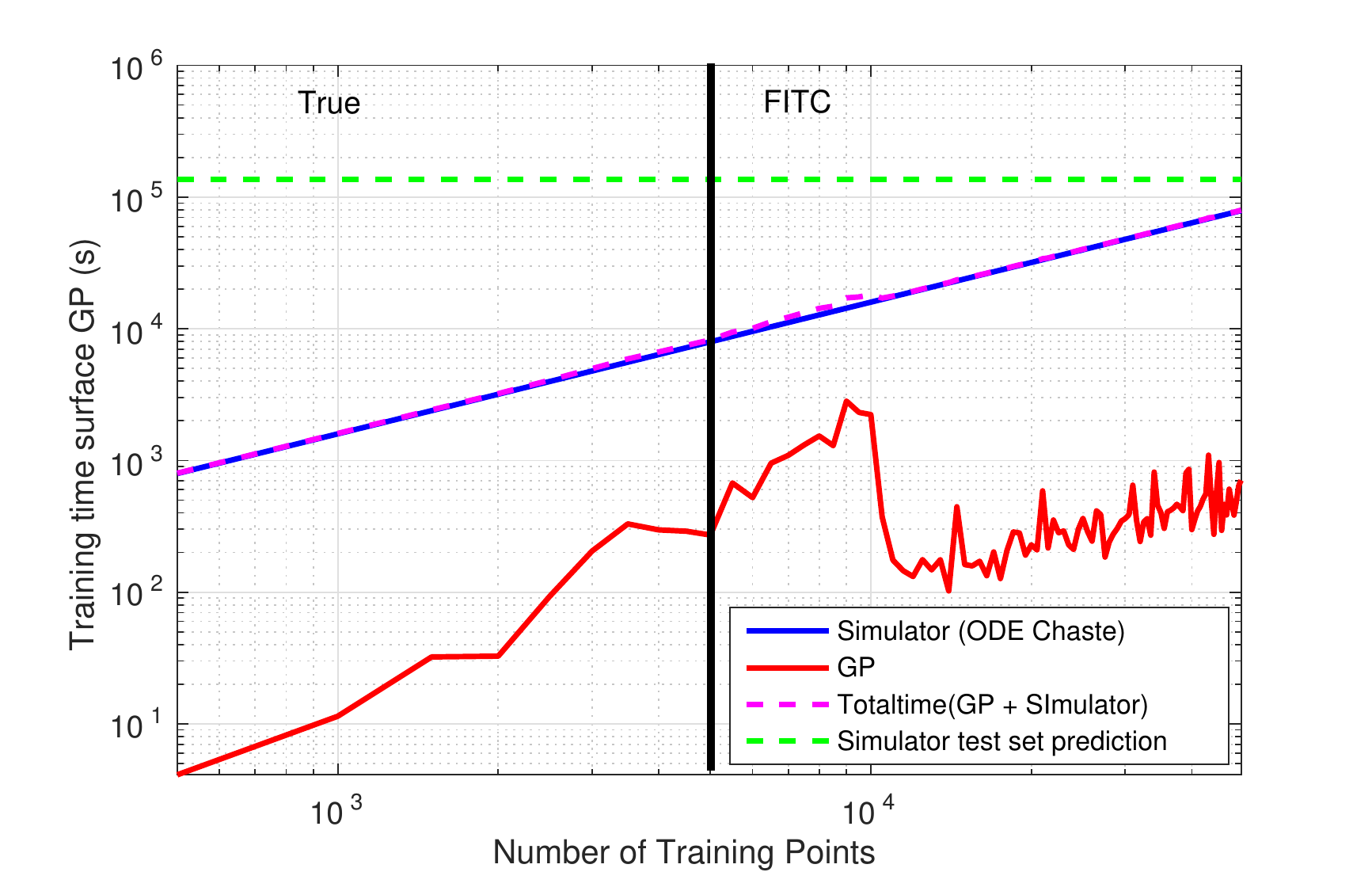}
    \label{Figure:surfPredInterpVsGplog}
  }  
	\subfigure[]{
    \includegraphics[width=5in,height=3in]{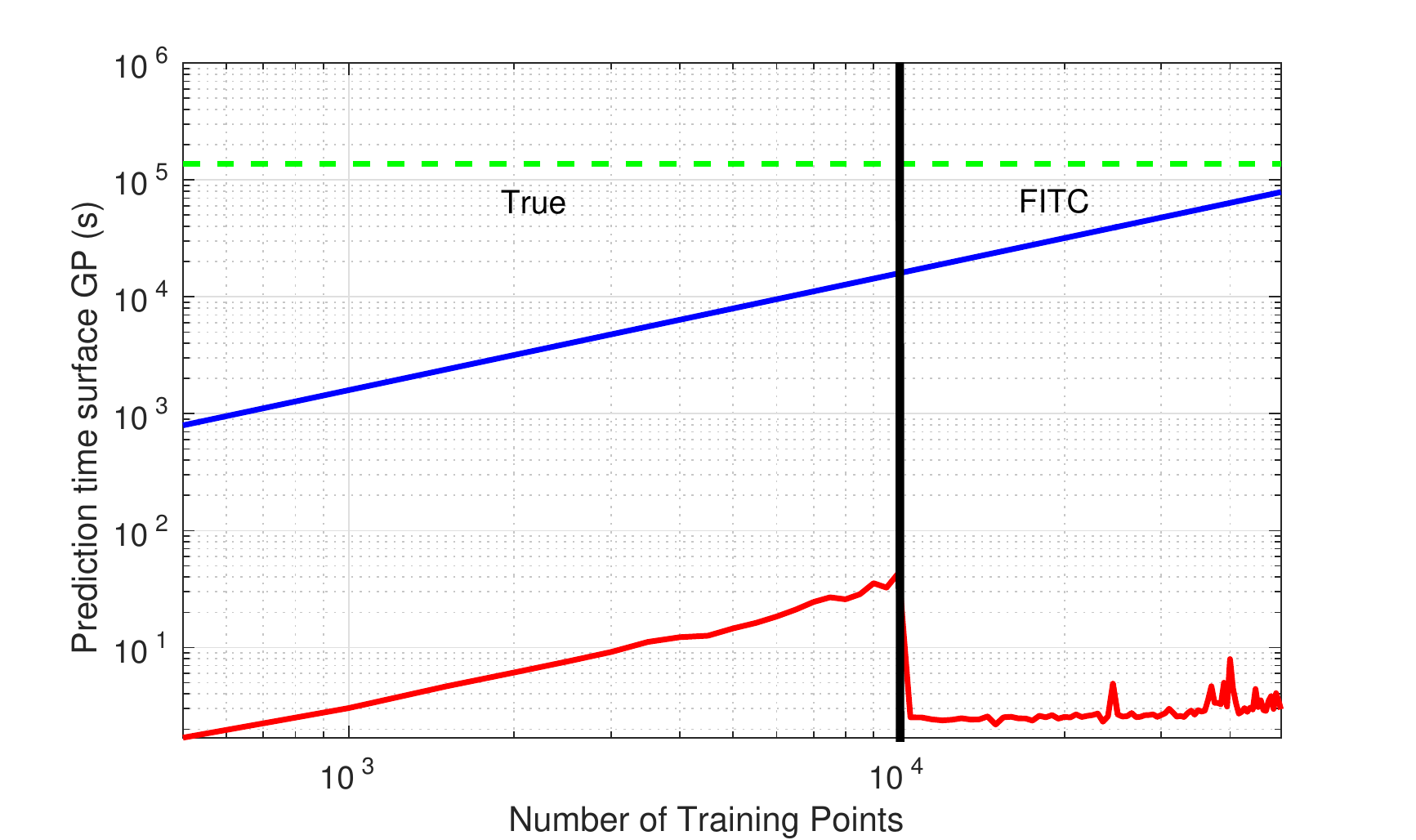}
    \label{Figure:surfTrainInterpVsGplog}
  } 
  \caption{\textbf{Surface GP timing performance for the 4-input problem, see section \ref{sec:Surface and boundary learning curves: random method} for the learning curves.} a) Training, and b) Prediction times with increasing numbers of training inputs. The predictions are drawn over the test dataset $\mathcal{D}_{test}$ which contains $100,000$ test points. The green broken line shows the time required by the simulator to evaluate $\mathcal{D}_{test}$. The blue line shows the simulation time for an increasing number of inputs and the red line shows the GP training (hyperparameter learning) and prediction time. The magenta line shows the total training time which is the sum total of the simulation and GP training time. The black vertical line demarcates the training size beyond which we use the FITC covariance.
  	Here we see the potential benefit in terms of speed when using the FITC method, despite the slightly larger error that we observed in Fig.~\ref{Figure:GPvsIntlsurfError}.
  }
  \label{Figure:GPvsIntlsurfTime}  
\end{figure*}

\begin{figure*}[!htb]
  \centering
  \subfigure[]{
    \includegraphics[width=5in,height=3in]{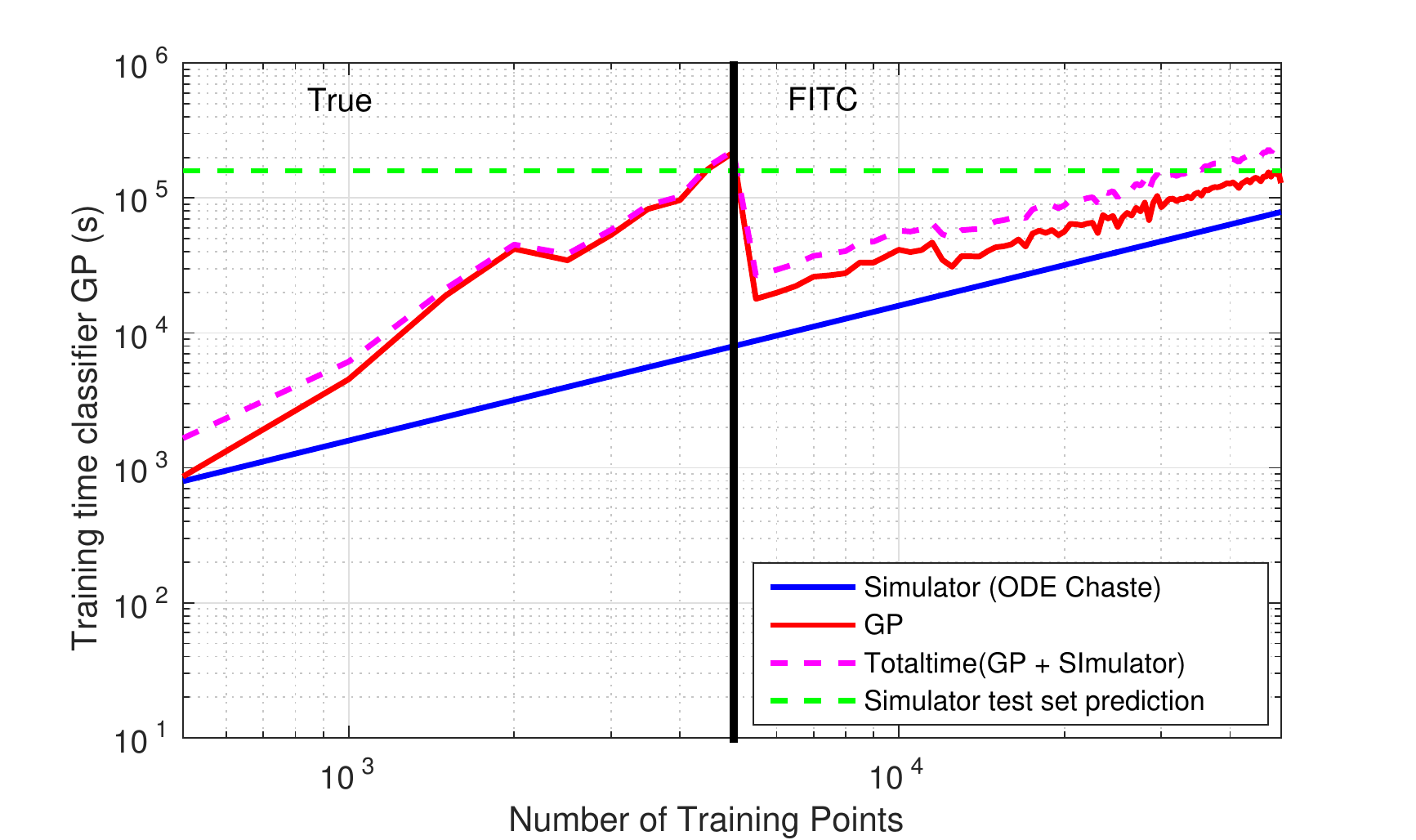}
    \label{Figure:classPredInterpVsGplog}
  }  
	\subfigure[]{
    \includegraphics[width=5in,height=3in]{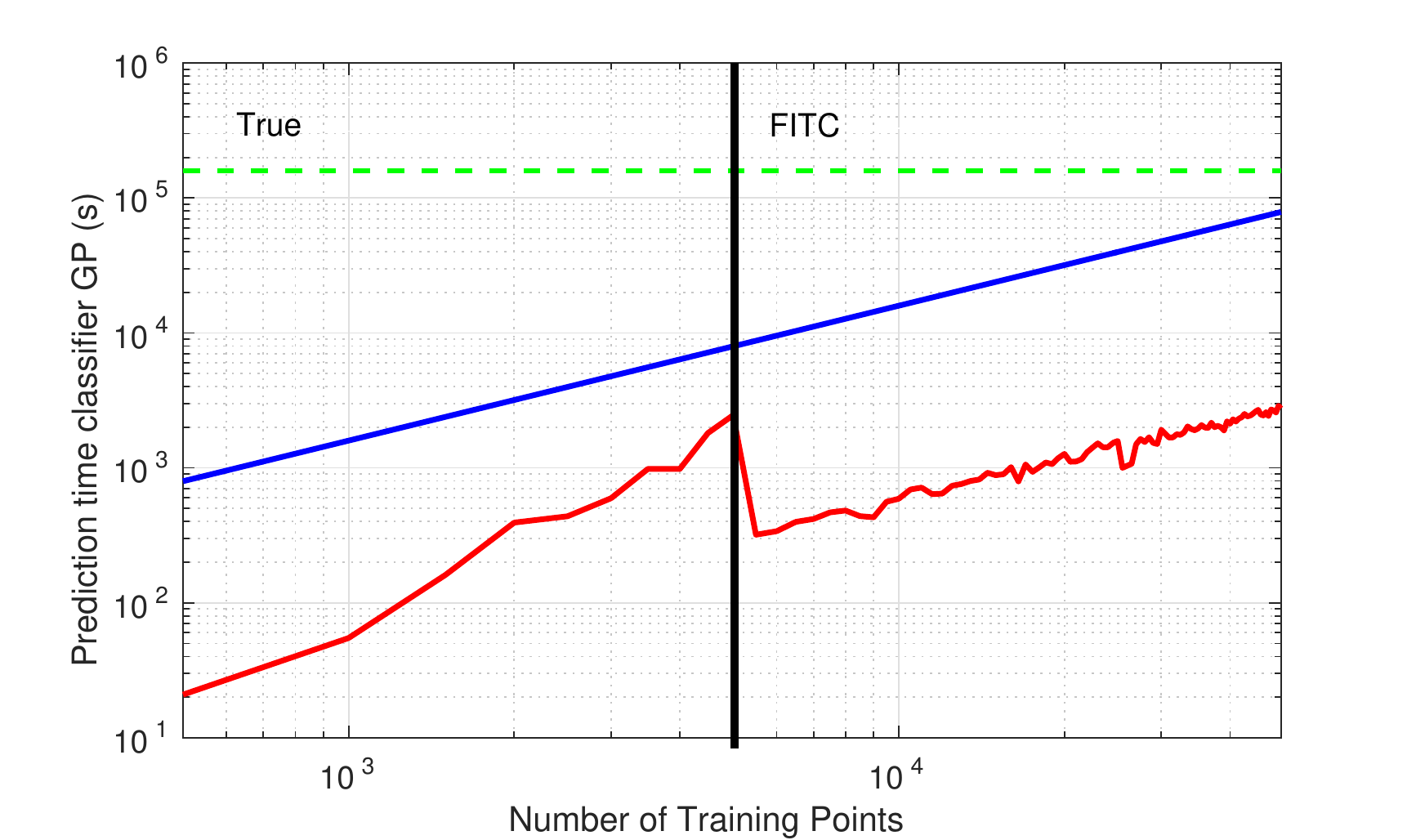}
    \label{Figure:classTrainInterpVsGplog}
  } 
  \caption{\textbf{ Training a) and Prediction b) time for the classifier GP for the 4-input problem, see section \ref{sec:Surface and boundary learning curves: random method} for the learning curves, with increasing number of training inputs}. The predictions are drawn over the test dataset $\mathcal{D}_{test}$ which contains $100,000$ test points. The green broken line shows the time required by the simulator to evaluate $\mathcal{D}_{test}$. The blue line shows the simulation time for increasing number of inputs and the red line shows the GP training (hyperparameter learning) and prediction time. The magenta line shows the total training time which is the sum total of the simulation and GP training time. The black vertical line demarcates the training size beyond which we use the FITC covariance.}
  \label{Figure:GPvsIntlclassTime}  
\end{figure*}

While designing the emulator we have to consider the fact that training and prediction of GPs are limited due to the computationally expensive covariance inversion step. 
For the classifier GP both training and prediction involve a number of such covariance inversion steps due to the expectation propagation algorithm. 
In Figs.~\ref{Figure:GPvsIntlsurfTime} \&  \ref{Figure:GPvsIntlclassTime} we show the training and prediction times for random learning in the case of the surface and classifier GP respectively. 
The corresponding learning curves are presented in Figs.~\ref{Figure:GPvsIntlsurfError}~\&~\ref{Figure:GPvsIntlclassError}. 
We have also plotted the simulation time (shown as a solid blue line) required to evaluate the corresponding number of training points for each of the GPs. 
The simulation time for evaluating all the points in the test set is shown as a horizontal broken line in all the plots. The total time for training the GP and simulating the training set is also plotted in Figs.~\ref{Figure:GPvsIntlsurfTime} \&  \ref{Figure:GPvsIntlclassTime}.

For the surface GP the total training plus prediction time is negligible in comparison to the simulation time for the entire test set, which is an expected result. 
Furthermore, using a FITC covariance we see further speed-up albeit at the cost of reduced accuracy (see Fig.~\ref{Figure:surfErrorInterpVsGplog} for the corresponding learning curve). 
However, for the classifier GP interestingly we see that while using a true covariance the training time exceeds the simulation time for evaluating the entire test set (mainly due to the PSO rounds). 
Even for the FITC covariance the training and prediction times are significantly higher than that of the surface GP. 
Thus in order to make our two-step approach work in a practical manner we need to use a small training dataset for the classifier GP. 
This is possible using the active learning scheme as we can use fewer points than the corresponding random learning.

\begin{figure}[!htb]
  \centering
  \includegraphics[width=5in,height=3in]{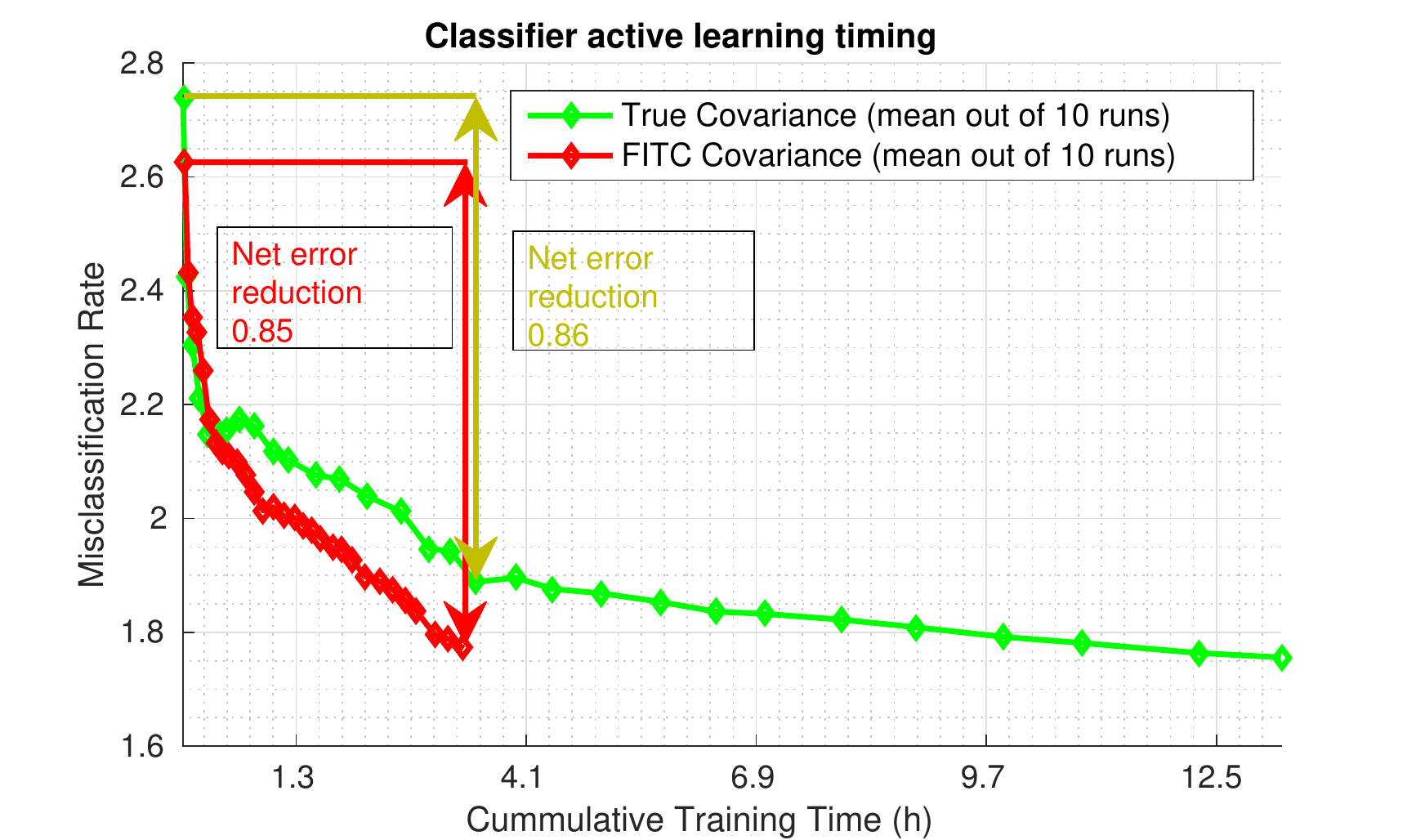}  
   \label{Figure:classTrueVsFitcTiming10unnorm}
   
  \caption{\textbf{Misclassification error reduces at the same rate using FITC and true covariance for carrying out active learning.} Comparison of true vs.\ FITC covariance for classifier active learning. 
  	Active learning is repeated $10$ times using different initial datasets. 
  	We notice same average (out of $10$ runs) rate of error reduction using both covariances. 
  	The horizontal and vertical lines point out the average error reduction observed after spending the time required for carrying out active learning using FITC (red lines). 
  	We used a swarm size of $n_s=50$ and a separate initial set of $n_1=500$ random training points for each repetition. 
  	Within this same time budget we achieve similar average error reduction using the true covariance (green line). }
  \label{Figure:classTrueVsFitcTiming}
\end{figure}

With a small number of training points we can reduce the training time of a classifier GP. 
But since each PSO round incurs many classifier predictions involving the EP algorithm it is important to find out the time spent in carrying out the classifier active learning, especially within the PSO iterations. 
In Fig.~\ref{Figure:classTrueVsFitcTiming} we plot the reduction in misclassification for the classifier active learning, averaged over $10$ repetitions, with increasing computation time. 
The computation time is represented as the cumulative training time which is the sum total of the training time upto the $r$-th round. 
We carried out the training for $30$ rounds with a swarm size of $n_s=50$. 
We also plot the same graph for the FITC covariance (red line). 
For each of the $10$ repetitions we used a separate initial set of $n_1=500$ random training points to learn the true and FITC covariance hyperparameters.

It is evident from Fig.~\ref{Figure:classTrueVsFitcTiming} that initially both using a FITC as well as the true covariance the average error is reduced almost at the same rate. 
However, we get marginally higher reduction using the FITC covariance. 

Another important algorithmic setting that has a potential impact on the run-time of the classifier active learning is the swarm size. 
A smaller swarm of points has to go through many more rounds of PSO in comparison to a larger swarm to collect the same amount of actively-learnt training points. 
However, more PSO rounds will enable the active learning scheme to hone in on different uncertain regions of the input space. 

\begin{figure}[!htb]
  \centering
  \includegraphics[width=5in,height=3in]{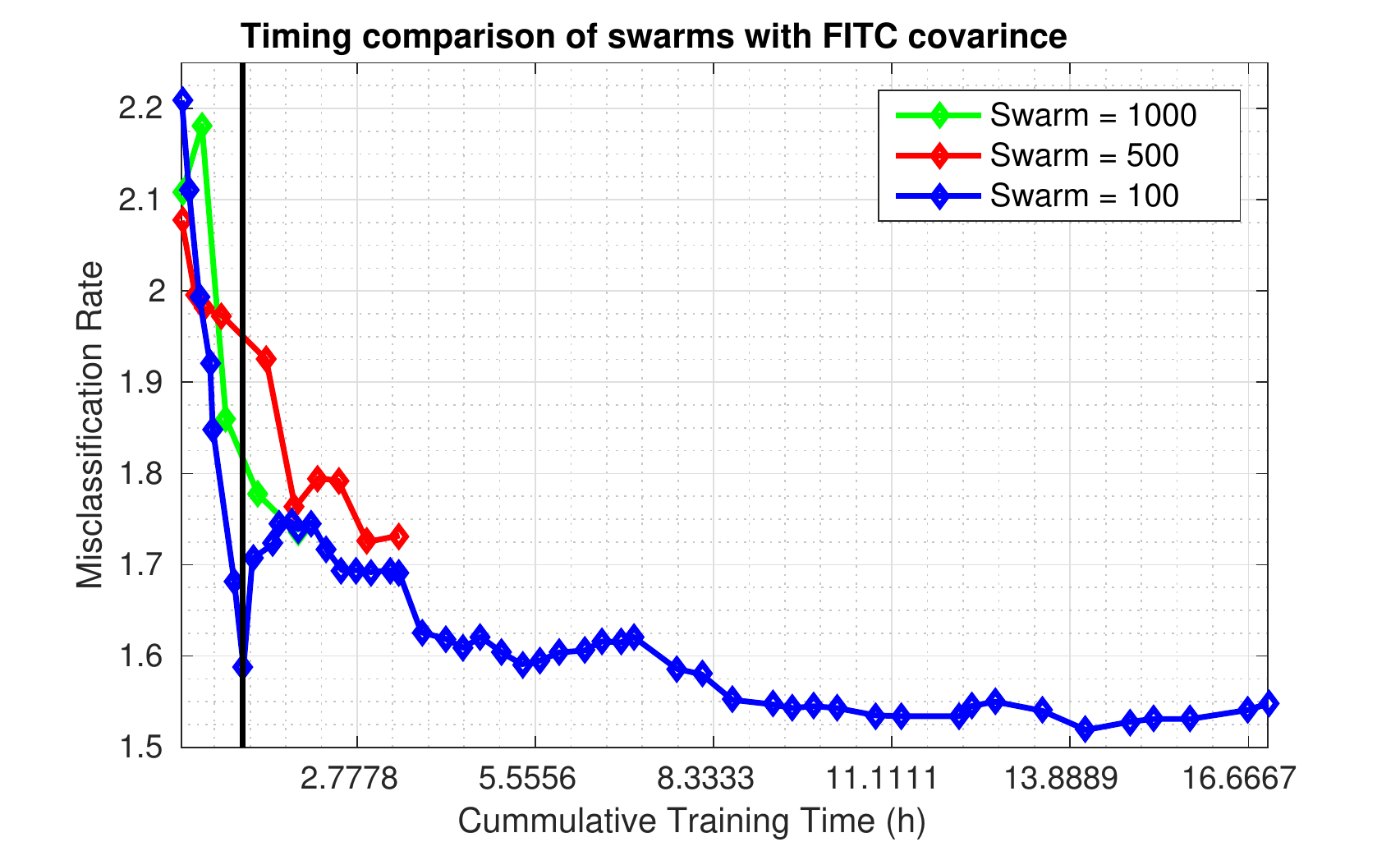}
  \caption{\textbf{More rounds of PSO with smaller swarm sizes reduces error most efficiently.} Effect of swarm sizes on active learning time with FITC covariance.
}
  \label{Figure:SwarmFitcTiming}
\end{figure}

In order to test this we ran the FITC active learning, with three different swarm sizes $\{1000,500,100\}$. 
We chose the initial dataset size $n_1$ to be $1000$ points. 
We carried out active learning for each of the swarm sizes to collect $5000$ active points. 
Hence, the GP with a swarm size of $100$ finished $l=50$ rounds while the GP with swarm size of $1000$ finished only $5 $ rounds. 
In Fig.~\ref{Figure:SwarmFitcTiming} we plot the misclassification rate for the three GPs concerned. 
The smallest swarm has the largest run-time but is able to reduce the error much more than the larger swarm variants. 
In fact the smallest swarm variant reduces the classifier error significantly more within $1$ hour (this is demarcated with the black line) than the larger variants. 
The reason for this is that the smallest swarm can complete more rounds of PSO within the same time compared to others and thus can put points across more distinct uncertain regions than the larger swarms do.
Covering more regions of uncertainty (with fewer points) is more important than covering fewer regions with more points.

\subsection{Performance of the two-step method} \label{sec:Performance of the two-step method}

Finally to test the performance of the complete two-step emulator we carry out active learning for both the classifier and surface GPs. 
For the classifier GP we used a swarm of $n_s=50$ points with $r=30$ rounds of PSO generating the active set of points $n_3=r \times n_s=1500$. 
We retain the same PSO threshold of $\theta=0.5$. 
We used the FITC covariance for the classifier GP. 

Subsequently for the surface GP we carry out active learning to gather $n_2=3000$ points sequentially. 
For the surface active learning we have used a candidate set of $10{,}000$ input points and used the actively learnt classifier to filter out all the non-AP points.
We chose an initial dataset $\mathcal{D}_{\emptyset}$ of size $n_1=500$, drawn randomly, to learn the classifier and surface GP hyperparameters. 
Thus, the total training data size for the two-step emulator is $n_1+ \hat{n}_2 + n_3=5000$.
As we filter out the non-AP points from the active surface training inputs our actual training size is slightly less than $5000$. 

To carry out a fair comparison between active and random learning the hyperparameters were learnt using $\mathcal{D}_{\emptyset}$ and were not updated after finishing all the rounds.
However, while evaluating the two-step emulator we have updated the hyperparameters after finishing the active learning. 
This update step affects the run-time of the two-step method and one can optionally avoid this step. 

In Table \ref{Table:paramST} we present the performance of the two-step GP which we denote as \textbf{Two-step} including the parameter re-learning/update step. 
We have also presented the performance of our previously described linear interpolator \cite{mirams2014prediction} denoted as the \textbf{Interpolator}, here trained using $5000$ points.
We also show the time required for both the GP based and interpolation method to complete the emulation task, which we denote as prediction time, and simulation (Chaste ODE evaluations) for the entire test dataset.
In Table~\ref{Table:UQ time} we breakdown the run-time of the emulator into training and prediction times for both the surface and boundary GP. 
In Table~\ref{Table:UQ time} the training time of classifier consists of the sum total of hyperparameter learning on $n_1=500$ initial points, active learning to collect $n_3=1500$ points and re-learning hyperparametrs on $2000$ points. 
For the surface GP the total training time consists of hyperparameter learning on valid AP points in the initial set of $n_1=500$, active learning to collect $n_2=3000$ points and re-learning hyperparametrs on the AP ones out of $3500$ points. 
Note that the surface and classifier GP training times include the time for ODE simulation during the respective active learning processes. 
The prediction is drawn on 100,000 test points.
\begin{table*}[!htb]
  \centering
	  \caption{\textbf{Performance of two-step emulator training for a $4$-input problem}. 
	  	We evaluated emulator performance by comparing against a test dataset of 100,000 points, picked from the 4D input space at random. 
	  	Simulating the full ODE solutions for these points took $44.2$ hours. 
	  	The training times for both the methods are listed below and include walltime for ODE simulation at the training points. 
	  	These performance tests were carried out on single processor (3GHz with 32GB RAM). 
	  	Although the two-step emulator's training time is double that of the interpolator it outperforms the interpolator in terms of prediction accuracies.
      }
  \label{Table:paramST}
  \resizebox{\textwidth}{!}{
  \begin{tabular}{l*{5}{c}r}
	\toprule
 Method & Training size & Training time (h)& Prediction time (s) & $E_{boundary}$ (\%) & $E_{surface}$ (ms)\\
\midrule
$\textbf{Two-step-GP}$ & $5000$ & $5.5085$ & $68.8980$ & $1.5770$ & $2.8742$ \\

$\textbf{Interpolator}$ & $5000$ & $2.4320$ & $1.8951$ & $13.9670$ & $17.9525$ \\

\bottomrule
\end{tabular}}

\end{table*}
\begin{table*}
  \centering
	  \caption{\textbf{Breakdown of run-time of the two-step emulator}. 
      Here the training time of the classifier and surface GPs constitute the sum total of hyperparameter learning on the initial and final training set as well as the active learning. 
      The total number of training points for the classifier is $n_1+(n_s \times r)=500+(30\times 50)=2000$ and for the surface is $n_1+n_2=500+3000=3500$.  %
      Predictions are made for a test-set of 100,000 points.}
  \label{Table:UQ time}
  \begin{tabular}{l*{5}{c}r}
	\toprule
 Method & Training time (h) & Prediction time (minutes)\\
\midrule
$\textbf{Classifier GP}$ & $4.2939$ & $0.6320$ \\
$\textbf{Surface GP}$ & $1.2146$ & $0.5163$ \\

\bottomrule
\end{tabular}
\end{table*}

Although the new two-step GP emulator is slower than the Look Up Table-based interpolator, considering the time needed in simulating the entire test dataset it is a reasonable alternative to the interpolator due to its improved accuracy in the presence of discontinuities. 
Also note that once the emulator is trained we can use it to evaluate the response surface for a large number of inputs repeatedly. 
In such a scenario the fast prediction time, as seen in Table~\ref{Table:UQ time}, is extremely valuable for an uncertainty propagation task.

\subsection{Drug Block Case Study} \label{sec:DrugBlock}

Finally, we evaluate the performance of the two-step emulator within a study of drug action. 
This is relevant to the work being undertaken in the Comprehensive in-vitro Pro-arryhthmia Assay initative \cite{fermini2016new} and will make the Uncertainty Quantification undertaken there \cite{chang2017uncertainty} faster to perform. 
Recently, Crumb \emph{et al.}\ \cite{CRUMB2016251} published dose-response screening data for $30$ compounds on $7$ different ion channels along with point estimates of $\left[IC50 \right]$ and the Hill coefficient $n$. 
Johnstone \emph{et al.}\ \cite{rossHill} then implemented a method to derive a probability distribution for the drug block parameters, as given by equation (\ref{eq:ce curve}), on various ion-channels. 
To propagate the uncertainty, as captured through the marginal posterior distributions of these parameters, APD\textsubscript{90} values were simulated using a Monte Carlo method for the corresponding samples of $\left[IC50 \right]$ and $n$.
We will test our emulator in the same setting, to establish whether it can provide the same insights in a more computationally tractable fashion.

\begin{figure}[!htb]
  \centering
  \includegraphics[width=0.9\textwidth]{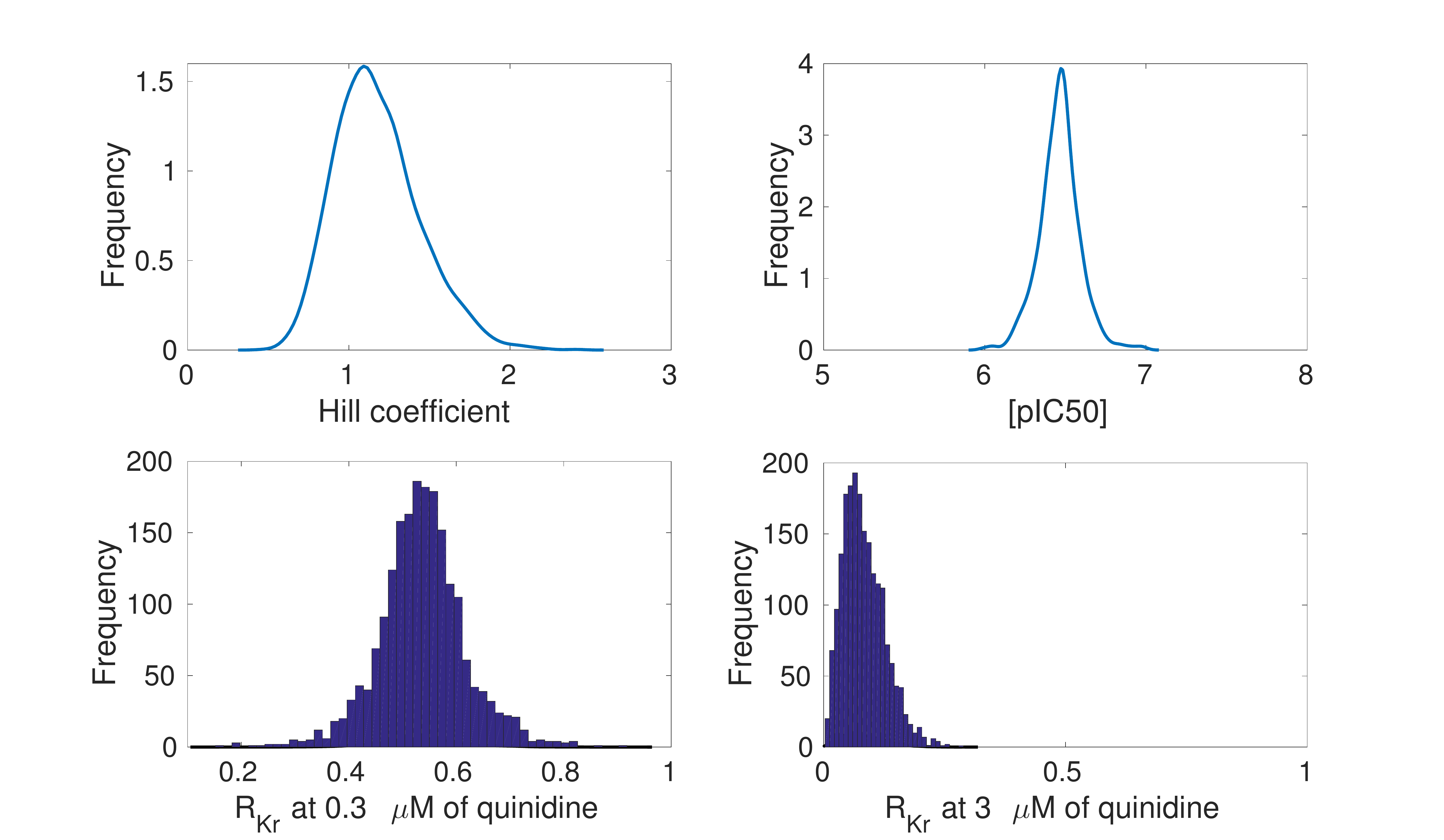}
  \caption{\textbf{Distributions of the concentration-effect-curve parameters as obtained in \cite{rossHill} through MCMC, and the corresponding distribution of $\Rb_{Kr}$ (hERG ion channel scalings) at different doses of quinidine.}   
  The top row shows the marginal posterior distributions (as kernel density estimates) of Hill coefficient $n$ and $\left[pIC50 \right]=-\log_{10}(\left[IC50 \right])$ estimated from the dose-response data in \cite{CRUMB2016251} for quinidine compound action on hERG channel. 
  The bottom row shows the corresponding distributions of $\Rb_{Kr}$, as histograms, at quinidine concentrations $0.3\,\mu$M (left) and $3\,\mu$M (right) calculated using equation (\ref{eq:ce curve}).   
  Each of the kernel density estimates and histograms are made using $2000$ samples.}
  \label{Figure:picHill}
\end{figure}

At a high concentration ($\left[C \right]$ in equation (\ref{eq:ce curve})) of quinidine, the reduction in hERG ion channel conductance pushes the model into the non-repolarising region. 
Thus, to evaluate our proposed emulator near to the discontinuous regime we repeat the uncertainty quantification task while blocking the hERG channel based on quinidine as the chosen drug.
We refer the reader to Johnstone \emph{et al.}\ \cite{rossHill} (section 4 in particular) for further details of the characterisation of input uncertainties, and have generated samples of the $\left[IC50 \right]$ and Hill coefficient $n$ using the technique and code they provided.
We used the hierarchical model variant from that paper and generated $2000$ samples inferring the underlying drug effect (rather than including a prediction of future experiment-level variability in our samples) using the concentration effect curve given by equation (\ref{eq:ce curve}) for 
i) a moderate dose --- $0.3\,\mu$M of quinidine producing $\approx 50\%$ block and 
ii) a higher dose --- $3\,\mu$M of quinidine producing $\approx 85\%$ block, obtaining distributions of conductance scalings $\Rb_{Kr}$ shown in Fig. \ref{Figure:picHill}. 
We picked these concentrations to test the emulator for uncertainty quantification on a distribution of APD\textsubscript{90} that is i) entirely on the surface and ii) straddling a discontinuity.
Our distributions shown in Fig.~\ref{Figure:picHill} are analogous to those shown in the original publication \cite{rossHill} (Fig. 12E in that paper).

Since in many drug action studies we would be testing a single ion channel, the conductance for that channel would be blocked, whereas for other channels the conductances would be set to the maximal conductance with no blocking. 
To account for this scenario we augment the initial random dataset $\mathcal{D}_{\emptyset}$ with $2^4$ training points which have the scalings $\Rb$ set to $0/1$, and use a full 4D emulator.
With this new augmented $\mathcal{D}_{\emptyset}$ we repeat the active learning for both the classifier and surface GP as described in the previous section. 
The total training set size for the two-step emulator in this case is $n_1 + 2^4+ n_2 + n_3=5016$. 
Addition of these corner points does not change the prediction accuracy noticeably for a general case where we draw test points which have some amount of scaling applied to each channel such as our test dataset $\mathcal{D}_{test}$ used so far for testing the two-step emulator in the previous sections. 
Drawing predictions on $\mathcal{D}_{test}$ we found the surface error $E_{surface}=3.0597$ ms and the classifier error $E_{boundary}=1.5580\%$. 
We notice little difference between these error values and the ones reported in Table \ref{Table:paramST}.

\begin{figure}[!htb]
  \centering
  \includegraphics[width=0.85\textwidth]{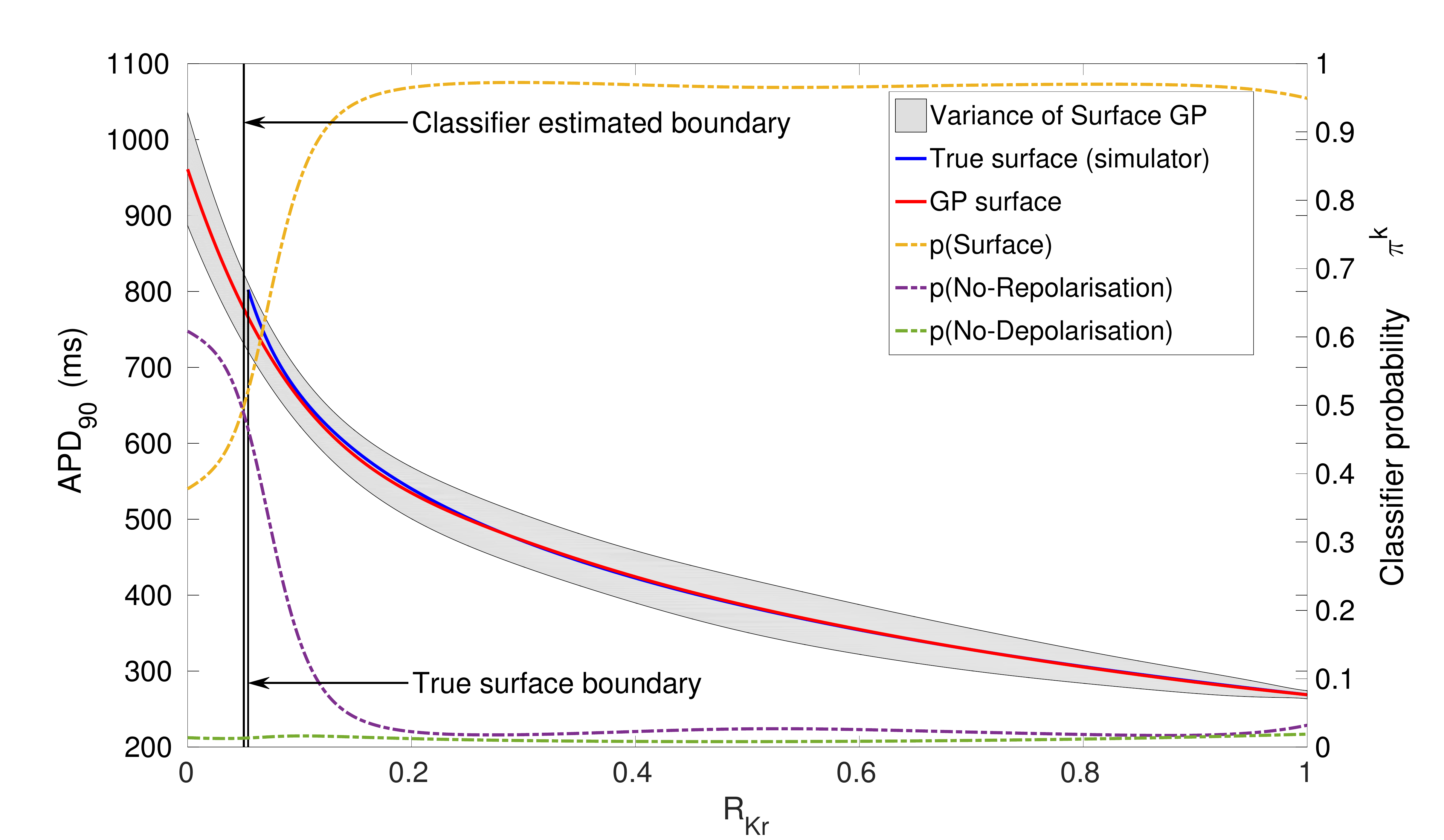}
  \caption{\textbf{$1$-dimensional slice of surface prediction along with binary classification posterior probabilities on a linear grid of $\Rb_{Kr}$. Misclassification is associated with low classifier probabilities.} 
  The surface GP prediction of APD\textsubscript{90} values is plotted (red line) along with the emulator variance (shaded area), scale shown on left hand axis.
  The true solution from running the ODE solver (simulator) is plotted as a blue line. 
  The true and the classifier's estimated class boundary are shown as vertical lines. 
  The posterior class probability $p$(Surface) of the surface (valid AP vs rest of the regions) region is shown on the right hand axis (orange line) and reduces rapidly in the misclassification region between the two vertical lines. 
  The variance of the surface GP also reduces in this region. 
  The corresponding posterior class probabilities $p$(No-Repolarisation) and $p$(No-Depolarisation) of the non-repolarising and non-depolarising regions vs rest of the regions respectively, are also shown on the right hand axis as broken lines. 
  Note that the true solution is contained within the emulator variance.
}
  \label{Figure:1D-slice}
\end{figure}

We perform a prediction for all the 2000 samples of $\Rb_{Kr}$ at each of the two concentrations using the emulator to obtain estimates of the APD\textsubscript{90} surface using equation~(\ref{eq:APD pred}).
To facilitate the visualisation of the classifier and surface GP predictions we plot one-dimensional slices, in Figure~\ref{Figure:1D-slice}, of the pre-trained classifier posterior probabilities and the surface GP mean and variance  for an artificial test dataset with $1000$ samples of $\Rb_{Kr}$ spread evenly between $0$ and $1$. The posterior probabilities consist of the probabilities $\pi^k$ of the binary GP classifiers for all the three regions.
Note that for visualisation we are passing all the artificial test points to the surface GP, unlike the two step method where we pass only those which are classified as valid AP points to the surface GP. 
We set the scaling of other channels as: $\left(\Rb_{Na},\Rb_{Ks},\Rb_{CaL}\right)=1$, to represent no block at those channels. Whilst this means a 1D emulator could be used, we wish to test our more general 4D emulator in what follows.

In previous sections we considered the posterior mean of the surface GP at test points to define APD\textsubscript{90} surface. 
However, in a real application such as this drug action study we also want to include uncertainty due to the emulator, as estimated by its posterior variance; and we wish to propagate this uncertainty into the corresponding APD\textsubscript{90} distribution too.
 
Thus to calculate the full uncertainty we simply sum the continuous Gaussian distributions for APD given by the emulator at each discrete sample of block, and re-normalise to produce a full probability distribution for APD. 
We denote the number of test points classified as being in the `full AP region' as $N^{AP}$, so
\begin{equation}\label{eq: APD-UQ}
p_{APD\textsubscript{90}}=\frac{1}{N^{AP}}\sum_{i=1}^{N^{AP}} \mathcal{N}\big(\mu(f_i),Var(f_i)\big),
\end{equation}
where $\mu(f_i)$ (equation (\ref{GP emulator cond mean})) and $Var(f_i)$ (equation (\ref{GP emulator cond var})) are the posterior mean and variance of the surface GP at the $i$-th (out of $N^{AP}$) test point. 
Computationally, we discretise the above continuous equation into $1000$ values of APD\textsubscript{90} between $0$ and $1000$\,ms (the bounds for a 1\,Hz simulation).
 
\begin{figure}[!htb]
  \centering
  \includegraphics[width=0.75\textwidth]{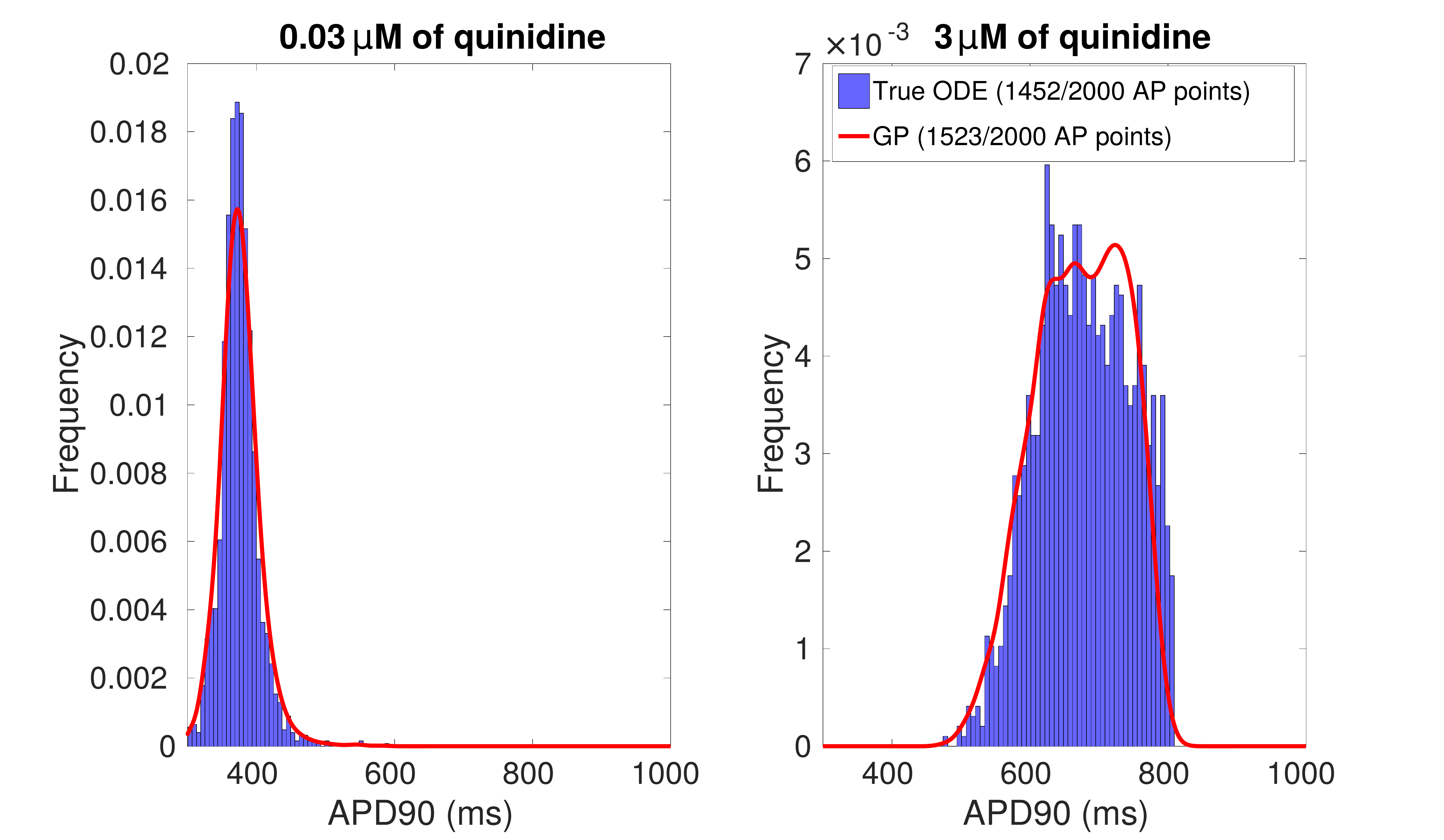}
  \caption{\textbf{Uncertainty propagation from concentration-effect-curve parameters to APD\textsubscript{90}.} 
  Distributions of APD\textsubscript{90} obtained through emulation (evaluated as given in equation (\ref{eq: APD-UQ})) are plotted as red lines for corresponding $\Rb_{Kr}$ values as represented in Fig.~\ref{Figure:picHill}. 
  Distributions of APD\textsubscript{90} obtained through Monte Carlo samples with a full simulation of the ODE system are shown as histograms. 
  }
  \label{Figure:APD-UQ}
\end{figure}

Fig.~\ref{Figure:APD-UQ} shows the distribution of APD\textsubscript{90} from taking samples from Fig.~\ref{Figure:picHill} and mapping them through the classifier and Surface GP shown in Fig.~\ref{Figure:1D-slice} and finally evaluating equation (\ref{eq: APD-UQ}). 
We also evaluate $p_{APD\textsubscript{90}}$ using the ODE simulator directly on all samples of block for comparison.

For both moderate and high dose cases we find that the distributions of APD\textsubscript{90} obtained from the emulator and simulator match well. 
For the moderate dose there is no misclassification. 
For the high dose $548$ out of the $2000$ $\Rb_{Kr}$ samples belong to the non-repolarising region when running the full ODE system. 
The emulator assigns $477$ points to this region and thus misclassifies $71$ points. 
The slight increase in misclassification rate $3.55\%$ from what is reported in Table \ref{Table:paramST} for predictions on $\mathcal{D}_{test}$ happens due to the presence of most of the test samples in this UQ task being near the class boundary.

We notice from Fig.~\ref{Figure:1D-slice} that the binary classifier probabilities for the AP and the non-repolarising regions are almost the same near the class boundary; as a result the certainty is very low in this region. 
If using the emulator for safety critical applications (such as the high dose quinidine action study considered here) if some test input points are located right along the discontinuities then one switch to performing full simulation simulation for points where the certainty $c$ is less than a threshold, say $0.8$.

\section{Discussion} \label{sec:Discussion}

Uncertainty and variability is intrinsic to a plethora of biological processes that we want to understand, model and predict. 
In cardiac modelling, sources of uncertainty stem from the experimental `error' in the measurements from our protocols, lack of knowledge about the underlying mechanisms leading to `structural error' in our models, variability due to differences in cell and ion channel states due to cells being in different settings and gene expression patterns, and variability due to the inherent stochasticity of some of these processes exhibited at multiple time and spatial scales \cite{mirams2016}. 
To accommodate mathematical/phenomenological models in safety-critical clinical practice and drug development, it is therefore of utmost importance to quantify and propagate these uncertainties to model predictions. 
As a consequence we need to examine our model predictions over large parameter domains. 
This is where emulation becomes necessary to reduce the computational burden associated with uncertainty quantification and propagation. 
However, many mathematical models, especially in cardiac electrophysiology, have bifurcations in behaviour as we move through parameter domains, rendering traditional Gaussian Process emulation infeasible. 
In this work we have addressed this specific issue of emulating cardiac action potential models having bifurcations in dynamics, and as a result, discontinuous output/response surfaces. 
We proposed a two-step emulator combining GP classification and regression to emulate the discontinuous action potential duration biomarker response surface and applied our method to a study of drug action.

Looking at the computational complexity of the GP classifier it is natural to ask whether a simpler classifier could be used for boundary detection.
To achieve a good degree of separation we have to use many more training points with a simple classifier such as a linear softmax classifier (this classifier can generate probabilistic predictions). 
Furthermore, using the probabilistic framework of the GP classifier we can quantify the uncertainty of the boundary locations, somewhat accounting for the fact that numerical errors become important close to the bifurcation point and so the notion of a somewhat random and probabilistic outcome here is helpful even though the ODE system is completely deterministic in this case.
We use this probabilistic property to build an active learning scheme which reduces the necessary training dataset size significantly. 
Using a complex boundary detector we become less sensitive to simulation errors and are able to use the simulator more sparsely. 

We tried an alternative GP classifier using the Laplace approximation \cite{rasmussen2010gaussian} which reduced the training and prediction times dramatically (results not shown). 
However, there was a significant drop in accuracy, and so we retained use of the EP algorithm.

If there is sufficient computing power available, the classifier certainty metric could be used to confine use of the emulator to locations that we are confident in the class. 
Thus, we can associate a threshold, say $0.9$, and if for a particular test point the certainty is below the threshold then we can use simulation for finding the true output value.
We did not explore an adaptive train-use-refine scheme here, but it would then be intuitive to add the simulation points to the GP training sets.

We have so far not discussed the timing implications of the surface active learning. 
This is because the surface active learning took less than $30$ minutes (for $n_2=3000$) to finish. 
This speed can be further reduced by using the FITC approximation, but we recommend the usage of true covariance as the run-time is insignificant in comparison to classifier active learning.

In this paper we have confined our emulation to one biomarker: the APD. 
However, the effect of bifurcation is observed in many other summary statistics of the action potential too, and the techniques are transferable. 
Using multiple GPs one can build a two-step emulator that covers the output space consisting of all the relevant biomarkers of the action potential trace. Or, using a multi-output GP \cite{bonilla2007multi,durichen2015multitask}, correlations among the generated biomarkers can also be captured to enhance prediction accuracies.

\section{Conclusion} \label{sec:Conclusion}

In this paper we presented a two-step emulator of the discontinuous APD\textsubscript{90} biomarker response surface generated by the O'Hara cardiac AP model under varying drug block. 
The proposed emulator produces good prediction accuracies on an artificial test dataset containing 100,000 test points. 
Our two-step emulator was trained at a fraction \textbf{($\approx 10 \%$)} of the computational expense of simulation. 
The proposed emulator requires $\approx 1$ minute for drawing predictions on the entire test dataset.
We achieve this by using sparse GP approximation (FITC) in conjunction with a novel active learning scheme.
A significant amount of the emulation time is consumed by the classifier GP due to its inherent computational limitations stemming from repeated covariance inversion within the EP algorithm. 

We have applied our two-step method for uncertainty quantification in a drug action study. 
In this application we found the two-step method was able to perform uncertainty quantification of APD\textsubscript{90} with high prediction accuracy. 
Our proposed method can be easily extended to accommodate other biophysical models (that go through bifurcations) and biomarkers.

\section{Materials and Methods}\label{sec:materials}
All the codes we used to generate the results in this study are available to download from \texttt{https://github.com/sanmitraghosh/ApPredict\_GP}.

\section{Acknowledgements} \label{sec:Acknowledgements}
SG and DJG were funded through the UK Engineering and Physical Sciences Research Council 2020 Science programme (grant number EP/I017909/1).
SG and GRM acknowledge support from a Sir Henry Dale Fellowship to GRM jointly funded by the Wellcome Trust and the Royal Society (grant number 101222/Z/13/Z).

\bibliographystyle{ieeetr}
\bibliography{minths.bib}

\end{document}